\documentclass[12pt]{article}

\usepackage{newtxtext,newtxmath}

\usepackage{graphicx}
\usepackage{setspace}
\usepackage{listings}
\usepackage[letterpaper,margin=1in]{geometry}
\renewenvironment{abstract}{\quotation}{\endquotation}
\date{}

\makeatletter
\renewcommand{\fnum@figure}{\textbf{Figure \thefigure}}
\renewcommand{\fnum@table}{\textbf{Table \thetable}}
\usepackage{makecell}

\newcommand{\reminder}[1]{\textcolor{red}{#1}}

\makeatother

\usepackage{science/scicite}

\usepackage{url}

\providecommand{\citep}[1]{\cite{#1}}
\providecommand{\citet}[1]{\cite{#1}}

\usepackage{booktabs}
\usepackage{siunitx}
\usepackage{tabularray}
\UseTblrLibrary{booktabs,siunitx}

\usepackage{tikz}
\usepackage{subcaption}

\usetikzlibrary{calc,positioning,arrows.meta,patterns}

\providecommand{\reminder}[1]{}

\def\scititle{Revealing AI Reasoning Increases Trust but Crowds Out Unique Human Knowledge
}
\title{\bfseries \boldmath \scititle}

\author{%
  Zenan~Chen$^{1\dagger}$, Ruijiang~Gao$^{1\dagger}$, Yingzhi~Liang$^{1\dagger\ast}$\\[4pt]
  \small$^{1}$Naveen Jindal School of Management, University of Texas at Dallas, Richardson, TX, USA.\\[2pt]
  \small$^{\dagger}$Authors listed in alphabetical order.\\
  \small$^{\ast}$Corresponding author.\\[2pt]
  \small Emails: \texttt{\{zenan.chen,ruijiang.gao,yingzhi\}@utdallas.edu}
}

\begin{document}

\maketitle

\begin{abstract}

Effective human-AI collaboration requires humans to accurately gauge AI capabilities and calibrate their trust accordingly. Humans often have context-dependent private information, referred to as Unique Human Knowledge (UHK), that is crucial for deciding whether to accept or override AI's recommendations. We examine how displaying AI reasoning affects trust and UHK utilization through a pre-registered, incentive-compatible experiment (N = 752). We find that revealing AI reasoning, whether brief or extensive, acts as a powerful persuasive heuristic that significantly increases trust and agreement with AI recommendations. Rather than helping participants appropriately calibrate their trust, this transparency induces over-trust that crowds out UHK utilization. Our results highlight the need for careful consideration when revealing AI reasoning and call for better information design in human-AI collaboration systems.

\end{abstract}

\section{Introduction}

Human-AI collaboration has become increasingly involved in making high-stakes decisions, such as clinical diagnostics \cite{Tu2025-hi, Goh2024-bp}, radiology reporting \cite{Huang2025-rb}, customer support \cite{Brynjolfsson2025-sg}, and even autonomous laboratory work \cite{Boiko2023-vf}, shaping outcomes in health, business, and scientific discovery. The performance of human-AI teams depends crucially on whether humans can accurately assess AI's capabilities and know when to accept or override AI recommendations. To make these decisions, humans rely on their Unique Human Knowledge (UHK) to evaluate AI recommendations. We define UHK as private, tacit, or context-specific information that AI systems cannot directly observe or reliably infer \citep{fugener2021will, fugener2022cognitive, gao2025confounding,balakrishnan2025human}. When conflicts arise between AI recommendations and UHK, humans must effectively override AI to achieve optimal human-AI team performance.

However, humans struggle to calibrate their trust in AI. A growing literature documents \emph{algorithm appreciation} and \emph{overreliance}: people tend to defer to AI even when it is suboptimal or wrong, due to the persuasive force of machine recommendations and anchoring on initial AI outputs \citep{Logg2019-ul, chen2024large, dietvorst2015algorithm, Nguyen2024-fx}. Such overreliance can crowd out the use of UHK, leading to mis-delegation and degraded performance \citep{fugener2021will, fugener2022cognitive, hemmer2022effectinformationasymmetryhumanai}. As such, much of the research in human-AI collaboration is therefore devoted to finding methods that effectively calibrate trust in AI.

Recent advances in LLM reasoning may offer a solution to this challenge.
LLM reasoning is widely adopted to improve model performance. Prompting an LLM to ``think step-by-step'' (i.e., Chain-of-Thoughts) enables it to decompose complex problems into intermediate steps, significantly boosting performance \citet{wei2022chain}. More advanced reasoning techniques used in Large Reasoning Models (LRMs) further improve performance through strategies such as exploring multiple reasoning paths, self-reflection, and reinforcement learning \citet{wang2022self,yao2023tree, guo2025deepseek}. Leading AI products such as DeepSeek R1 (\textsc{DeepSeek}), GPT-5 (\textsc{OpenAI}), Gemini 2.5 Pro (\textsc{Google}), and Claude Sonnet 4.5 (\textsc{Anthropic}) all employ varying extensiveness of reasoning and reveal this reasoning to users. 

Such reasoning reflects an LLM’s problem-solving process, comprising metacognitive monitoring and self-checks that naturally involve inconsistency and uncertainty cues. This distinguishes LLM reasoning from explainable AI (XAI) explanations \citep{kim2025fostering}. While XAI aims to help users understand a model’s internal processes, AI reasoning aims to improve AI performance \citep{guo2025deepseek} and is not designed to be persuasive \citep{chen2025reasoning}.

Showing LLM reasoning may help users better gauge AI accuracy and reduce overreliance on incorrect answers \citep{kim2024m,kim2025fostering}. This effect should be amplified particularly when users possess UHK to evaluate the reasoning. However, showing reasoning may also lead to cognitive biases. Classic persuasion literature shows that longer, more elaborate messages can increase heuristic acceptance \citep{petty1986-yr,chaiken1980heuristic}. Recent evidence shows that providing \emph{ex post} LLM justifications can hinder human oversight \citep{lane2024Narrative}. As such, it is unclear how revealing reasoning affects users' trust in AI and human-AI collaboration outcomes.

These tensions motivate our central question: Does exposing AI reasoning increase or reduce the utilization of unique human knowledge? We examine how varying the extensiveness of displayed AI reasoning—\textit{none}, \textit{brief}, and \textit{extensive}—affects two key outcomes: (i) user trust in AI, and (ii) performance gains from UHK utilization.

\section*{Methods}

We answer these questions by conducting a pre-registered\footnote{This study was pre-registered at \url{https://aspredicted.org/q2b2-9y9g.pdf}}, incentive-compatible online experiment on Prolific with a sample size of 752 participants. Participants act as hiring managers who screen candidate resumes using an LLM-based decision support system. Each resume contains ratings from 1 to 5 for three characteristics: \emph{Education}, \emph{Experience}, and \emph{Personality Fit}. The third characteristic, \emph{Personality Fit}, is the UHK in our design and represents information obtained through in-person interactions. Ground-truth hiring outcomes (hire/no-hire) are generated by a logistic model using these three features, with coefficients of 1, 1.2, and 1.5, respectively. These weights are calibrated so that when humans and the AI are provided with the same 10 historical hiring decisions, the AI’s accuracy is higher than humans when humans do not have UHK, and lower than humans when humans have UHK. This synthetic task also removes any real-world confounds.

Participants make predictions about whether a candidate should be hired based on 10 historical hiring decisions. For each candidate, participants make two decisions: an initial decision without the assistance of AI and a final decision after viewing the AI’s recommendation. Both decisions are incentivized; participants earn \$0.075 for each correct decision. This design ensures participants make an independent decision first, allowing us to rule out common mechanisms of the automation bias, where participants use AI explanations as a cognitive shortcut \citep{Yin2025-ig, lane2024Narrative, skitka1999does,goddard2012automation}.

We use a 3$\times$2 factorial between-within subject design. The between-subject treatment varies the displayed reasoning level: \emph{no} reasoning, \emph{brief} reasoning, and \emph{extensive} reasoning. Participants are randomly assigned to one reasoning level and are balanced across demographics, employment status, and their knowledge of and trust in AI (Table \ref{tab:descriptive_stats} in Appendix A.1). The within-subject treatment varies whether participants have UHK (\emph{Personality Fit}). In Phase I (the first 20 rounds), participants only have information on two characteristics: \emph{Education} and \emph{Experience}. In Phase II (the second 20 rounds), information about \emph{Personality Fit} is added. Importantly, in both phases, the AI does not have access to this feature. Before participants make Phase II decisions, we explicitly inform them that the AI lacks the third feature. This design rules out the lack of feature transparency as the source of any treatment effect, distinguishing our study from earlier research on feature transparency \citep{balakrishnan2025human}. To prevent learning effects from varying across reasoning levels, we do not provide participants with feedback on the correctness of their decisions between rounds.

To ensure that participants understand how to use historical hiring decisions for inference, they complete a training session before Phase I , in which they need to identify the hiring outcomes of three resumes drawn from the 10 historical cases. To emphasize the importance of UHK in Phase II, we then re-train participants using the same 10 historical data with the UHK (\emph{Personality Fit}) revealed. Figure \ref{fig:experimental_design} illustrates our experimental design.

\usetikzlibrary{positioning,calc,decorations.pathreplacing}

\begin{figure}[t]
\centering
\begingroup
\begin{singlespace*}
\resizebox{.95\linewidth}{!}{%
\begin{tikzpicture}[
    font=\small\sffamily,
    block/.style={draw, rounded corners, minimum width=2.6cm, minimum height=1.0cm, align=center, fill=white},
    label/.style={anchor=east, align=right},
    rowlab/.style={anchor=east, align=right},
    node distance=1.4cm and 0.95cm
]

\node[block] (t1) at (0,0) {Training};
\node[block, right=of t1] (n1) {No UHK (x20)};
\node[block, right=of n1] (rt1) {Re-Training};
\node[block, right=of rt1] (w1) {With UHK (x20)};

\node[block, below=0.6cm of t1] (t2) {Training};
\node[block, right=of t2] (n2) {No UHK (x20)};
\node[block, right=of n2] (rt2) {Re-Training};
\node[block, right=of rt2] (w2) {With UHK (x20)};

\node[block, below=0.6cm of t2] (t3) {Training};
\node[block, right=of t3] (n3) {No UHK (x20)};
\node[block, right=of n3] (rt3) {Re-Training};
\node[block, right=of rt3] (w3) {With UHK (x20)};

\node[rowlab] (r1label) at ($(t1.west)+(-0.30,0)$) {No reasoning};
\node[rowlab] (r2label) at ($(t2.west)+(-0.30,0)$) {Brief};
\node[rowlab] (r3label) at ($(t3.west)+(-0.30,0)$) {Extensive};

\foreach \a/\b in {t1/n1, n1/rt1, rt1/w1, t2/n2, n2/rt2, rt2/w2, t3/n3, n3/rt3, rt3/w3} {
  \draw[->, line width=1.0pt] (\a) -- (\b);
}

\draw[decorate, decoration={brace, amplitude=6pt, mirror, raise=2pt}]
  ($(t1.west)+(-3.4,0.15)$) -- ($(t3.west)+(-3.4,-0.15)$)
  node[midway, left=13pt, yshift=4pt, rotate=90, anchor=center] {\small Between-subject};

\draw[decorate, decoration={brace, amplitude=6pt, raise=2pt}]
  ($(t1.north west)+(0,1.0)$) -- ($(w1.north east)+(0,1.0)$)
  node[midway, yshift=15pt] {\small Within-subject};

\draw[decorate, decoration={brace, amplitude=5pt, raise=-5pt}]
  ($(t1.north west)+(0,0.5)$) -- ($(n1.north east)+(0,0.5)$)
  node[midway, yshift=10pt] {\small Phase I};

\draw[decorate, decoration={brace, amplitude=5pt, raise=-5pt}]
  ($(rt1.north west)+(0,0.5)$) -- ($(w1.north east)+(0,0.5)$)
  node[midway, yshift=10pt] {\small Phase II};

\end{tikzpicture}}
\end{singlespace*}
\endgroup
\vspace{5pt}
\caption{Experiment Procedure}
\label{fig:experimental_design}
\end{figure}
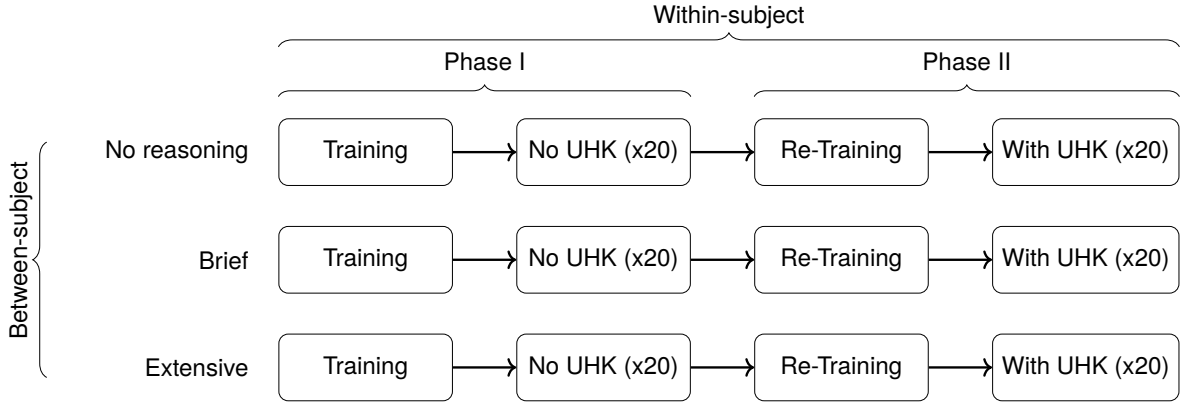

AI reasoning and recommendations are generated using a state-of-the-art reasoning \\LLM (\texttt{gemini-2.5-pro-preview}). To create different levels of reasoning extensiveness, we first extract the raw reasoning process from Gemini and create a streamlined version of this original reasoning, replicating the current practice of showing abbreviated paragraphs of reasoning processes used in ChatGPT, Gemini, etc. The \emph{extensive} reasoning condition preserves three different reasoning pathways that eventually reach the AI's final recommendation. The \emph{brief} reasoning condition is constructed by truncating the extensive reasoning, containing only the first reasoning pathway that arrive at the final recommendation. This design ensures that the brief reasoning condition is directly comparable to the extensive reasoning condition. In the \emph{no} reasoning condition, participants see a spinner with the text ``AI is thinking'' such that participants in all treatments understand the AI’s ability to reason, with the only difference being whether the reasoning is revealed. To ensure participants read the reasoning, all AI reasoning processes are presented through a word-by-word streaming animation. The exact prompts used to generate the AI recommendation and the original reasoning are provided in Appendix~\ref{app:prompts}. The prompt to generate our extensive and brief reasoning is provided in Appendix~\ref{app: reasoning_prompt} and an example of the different reasoning conditions is shown in Table~\ref{tab:reasoning_example} in Appendix~\ref{app: reasoning_prompt}.

Comparing user trust in AI across reasoning conditions within Phase I and within Phase II identifies the effect of reasoning on trust. Comparing decision accuracy between Phase I and Phase II identifies the effect of reasoning on UHK utilization. The complete descriptive statistics and balance checks are included in Table \ref{tab:descriptive_stats} in Appendix. 
\section*{Results}

\subsection*{Increased Trust in AI}

\begin{figure}[h!]
    \centering
    \begin{subfigure}[b]{0.48\textwidth}
        \centering
        \includegraphics[width=\textwidth]{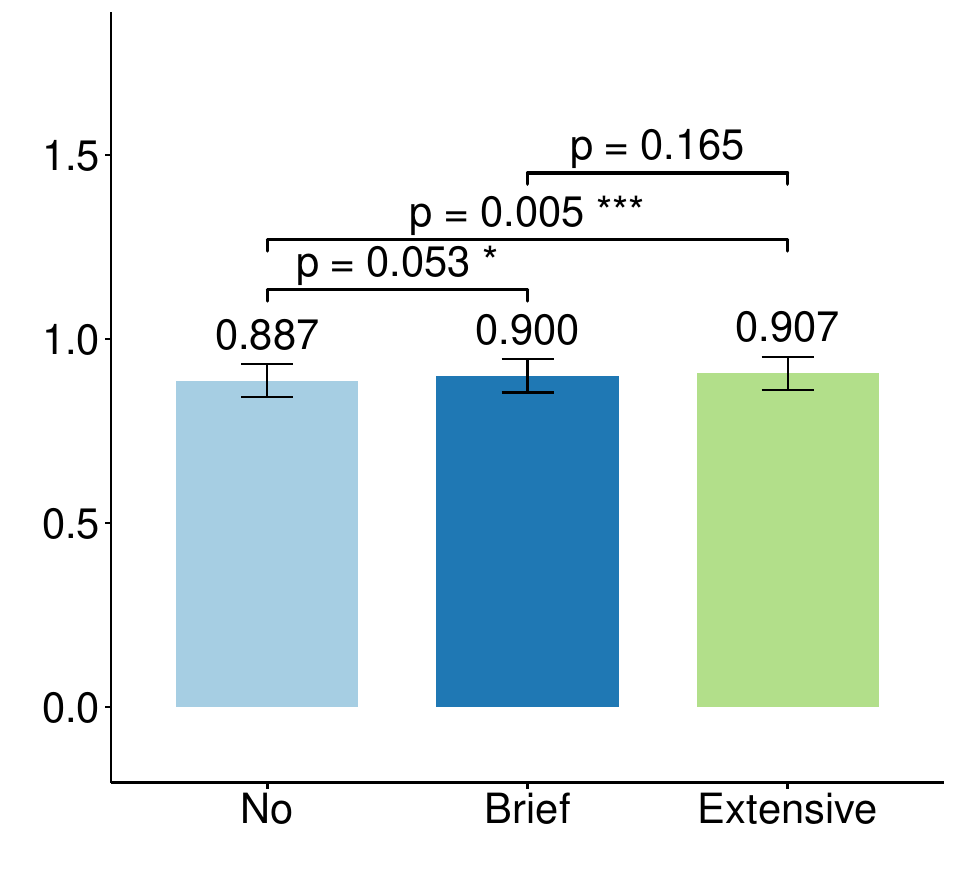}
        \caption{AI Agreement without UHK}
        \label{fig:no_uhk_agreement}
    \end{subfigure}
    \hfill
    \begin{subfigure}[b]{0.48\textwidth}
        \centering
        \includegraphics[width=\textwidth]{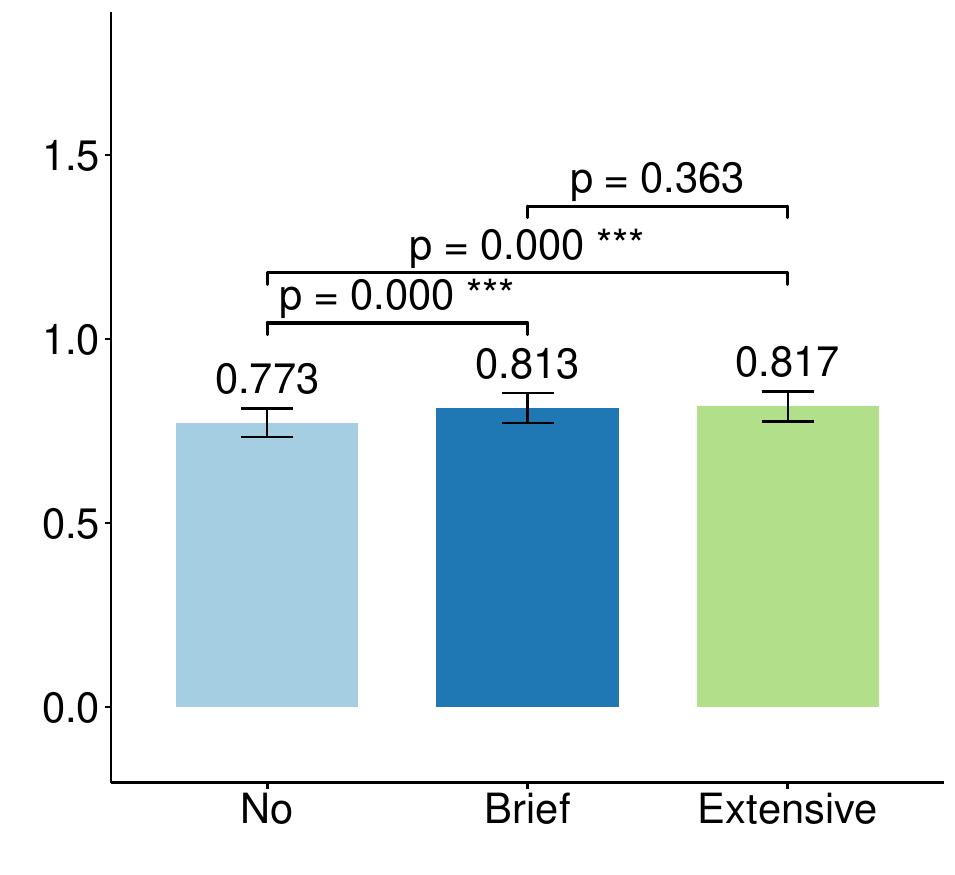}
        \caption{AI Agreement with UHK}
        \label{fig:uhk_agreement}
    \end{subfigure}
    \vskip\baselineskip
        \begin{subfigure}[b]{0.48\textwidth}
        \centering
        \includegraphics[width=\textwidth]{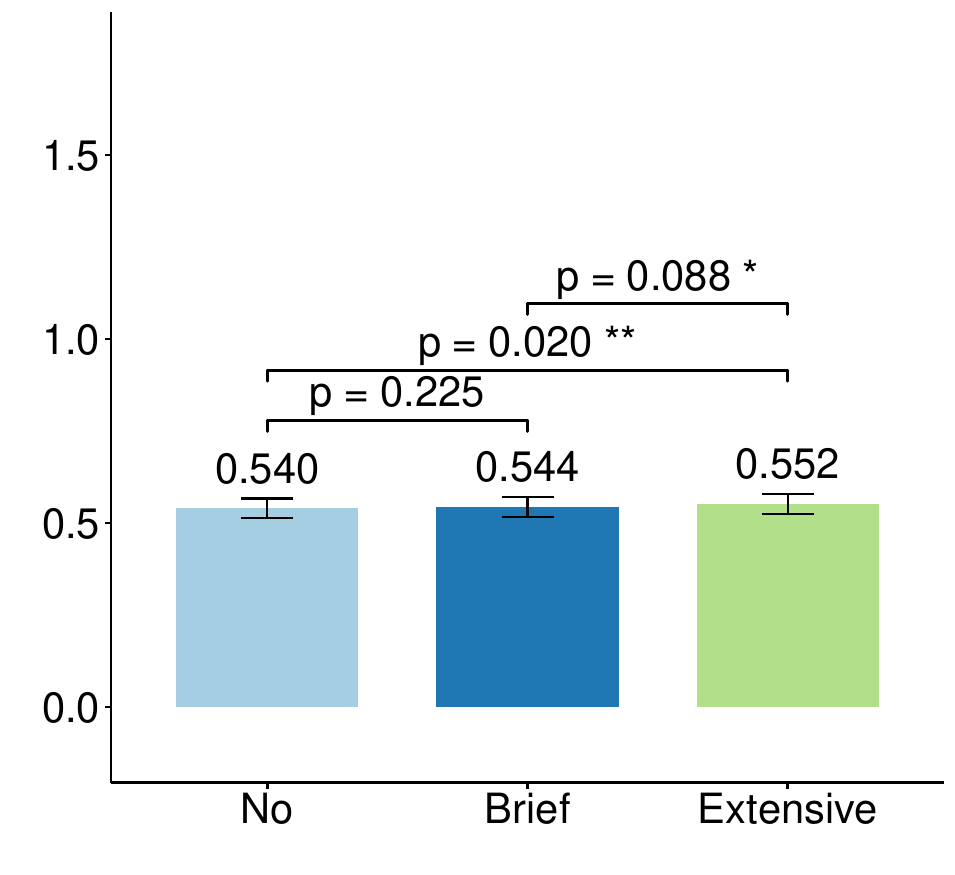}
        \caption{Accuracy without UHK}
        \label{fig:no_uhk_accuracy}
    \end{subfigure}
    \hfill 
    \begin{subfigure}[b]{0.48\textwidth}
        \centering
    \includegraphics[width=\textwidth]{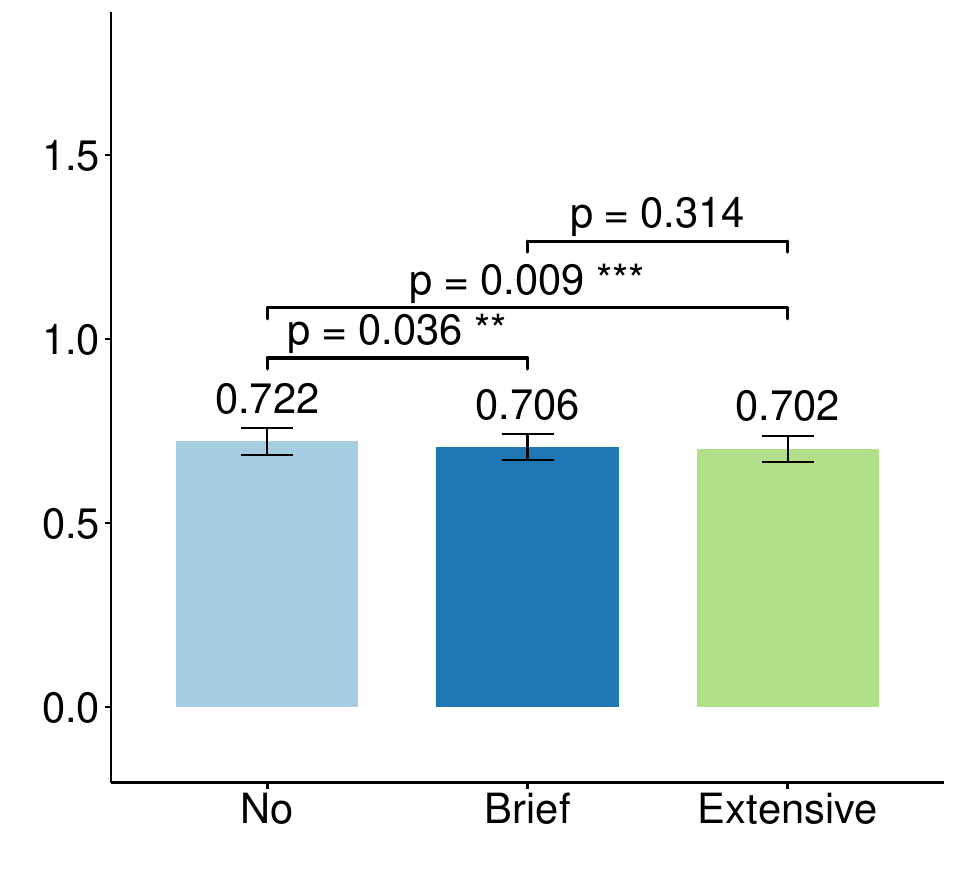}
        \caption{Accuracy with UHK}
        \label{fig:uhk_accuracy}
    \end{subfigure}
    \caption{Summary of accuracy and AI agreement.}
    \label{fig:performance}
\end{figure}

Displaying reasoning significantly increases users’ trust in AI, regardless of whether users possess UHK. Without UHK (Phase I), showing reasoning increases participant agreement with AI in their final decision from 88.7\% in the \emph{no} reasoning condition to 90.0\% with \emph{brief} reasoning ($p=0.053$) and 90.7\% with \emph{extensive} reasoning ($p=0.005$, Figure~\ref{fig:performance}a). With UHK present (Phase II), the same pattern holds, with agreement increasing from 77.3\% in the \emph{no} reasoning condition to 81.3\% with \emph{brief} reasoning ($p<0.001$) and 81.7\% with \emph{extensive} reasoning ($p<0.001$, Figure~\ref{fig:performance}b).

Our results show that even when participants were explicitly told that AI lacked the crucial \emph{Personality Fit} feature, they still trusted AI more after viewing its reasoning process. This indicates that rather than helping users calibrate when to defer, revealing the reasoning increased agreement with AI.

Surprisingly, the \emph{extensiveness} of the displayed reasoning did not affect user trust. AI agreement levels do not differ between the \emph{brief} and \emph{extensive} conditions, both without UHK and with UHK (See Figs. \ref{fig:no_uhk_agreement} and \ref{fig:uhk_agreement}). This contradicts with previous literature in persuasion which generally find that more elaborated arguments are more persuasive \citep{chaiken1980heuristic,petty1984effects}.

\begin{figure}[h!]
    \centering
    \begin{subfigure}[b]{0.48\textwidth}
        \centering
        \includegraphics[width=\textwidth]{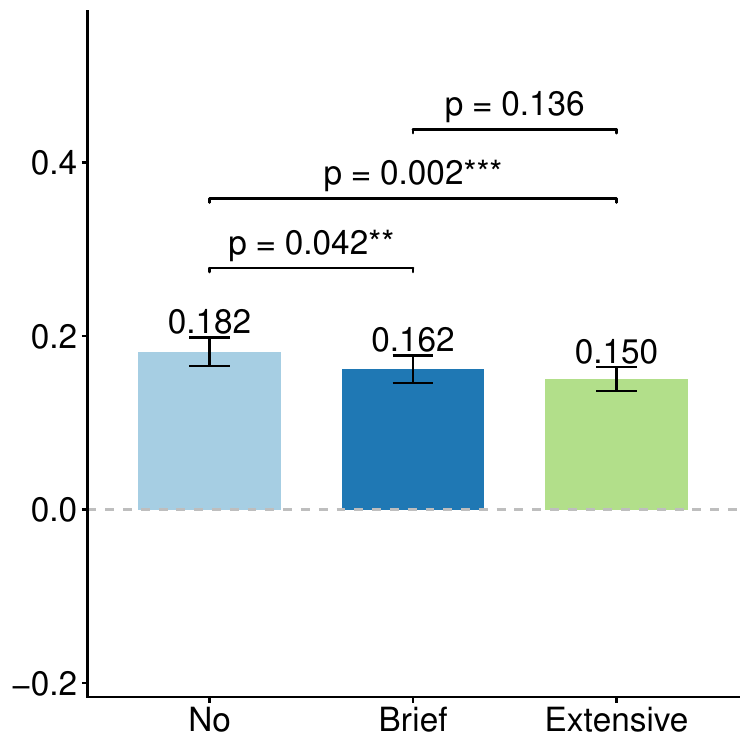}
        \caption{Change in Performance Gain}
        \label{fig:performance_change}
    \end{subfigure}
    \hfill
    \begin{subfigure}[b]{0.48\textwidth}
        \centering
        \includegraphics[width=\textwidth]{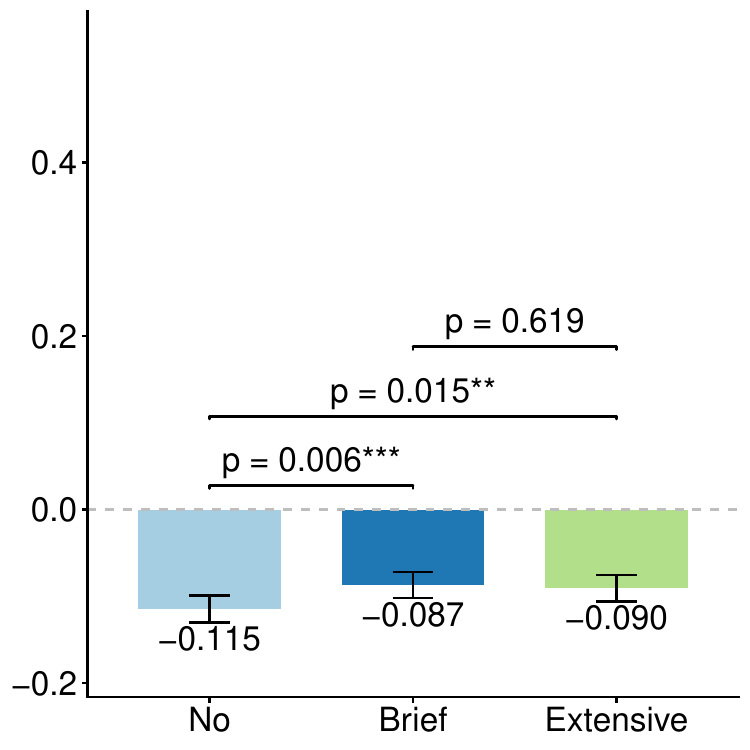}
        \caption{Change in AI Agreement}
        \label{fig:agreement_change}
    \end{subfigure}
    \caption{Phase II - Phase I}
    \label{fig: phase2_phase1}
\end{figure}

\subsection*{Performance Gain from UHK}

Across all reasoning conditions, final decision accuracy increases significantly when UHK is available (all $p<0.001$). The accuracy gain is 18.2\% in the \emph{no} reasoning group, 16.2\% in the \emph{brief} reasoning group, and 15.0\% in the \emph{extensive} reasoning group (see Figs. \ref{fig:no_uhk_accuracy}, \ref{fig:uhk_accuracy}, and \ref{fig:performance_change}).

At the same time, UHK reduces participants’ trust in the AI. AI agreement drops from Phase I to Phase II by 11.5\% in the \emph{no} reasoning condition, 8.7\% in the \emph{brief} reasoning condition, and 9.0\% in the \emph{extensive} reasoning condition (Fig.~\ref{fig:agreement_change}).

Taken together, these results show that across all reasoning conditions participants leverage their UHK to make better predictions, and they reduce their trust in the AI once they know it lacks the third feature.

\subsection*{Increased Trust Reduces Performance Gain}

Although access to UHK increases participants’ final prediction accuracy across all reasoning conditions, the magnitude of the gain differs. The \emph{no} reasoning condition shows the largest improvement (18.2\%), whereas gains in the \emph{brief} and \emph{extensive} conditions are significantly smaller: 16.2\% ($p=0.042$) in the \emph{brief} condition and 15.0\% ($p=0.002$) in the \emph{extensive} condition (Fig.~\ref{fig:performance_change}).

These reduced gains in the \emph{brief} and \emph{extensive} conditions appear to be driven by insufficient trust adjustments. While participants lower their trust in AI after receiving UHK under all conditions, the decline is smaller when reasoning is displayed. AI agreement rate falls by 11.5\% in the \emph{no} reasoning condition, compared with 8.7\% ($p=0.006$) in the \emph{brief} condition and 9.0\% ($p=0.015$) in the \emph{extensive} condition (Fig.~\ref{fig:agreement_change}).

This suggests that revealing the AI’s reasoning, regardless of extensiveness, biases participants’ judgments of the AI’s ability, even when they know it lacks a crucial feature, leading to over-trust in AI and under-utilization of their UHK. As a result, the performance gain from UHK is undermined.

\section*{Conclusions}

While LLM reasoning is an important method to improve LLM performance, our results show that exposing users to this reasoning can degrade their utilization of unique human knowledge.
In situations where humans and AI share the same information, showing reasoning leads to higher trust, which yields modest accuracy gains as users correctly lean on the AI's strengths. However, once users possess additional Unique Human Knowledge (UHK) that the AI lacks, this same trust becomes a liability. Instead of leveraging their exclusive insights, exposing reasoning makes users more likely to over-rely on AI, effectively crowding out their UHK.

Importantly, this degradation persists even when humans are explicitly told that the AI lacks key information and are required to make an independent decision before seeing the AI's recommendation. Our experimental design features rule out the lack of feature transparency and anchoring effects as explanations. Furthermore, the degradation cannot be mitigated by shortening the reasoning; our brief and extensive reasoning conditions generated similar effect sizes.

Our findings call for caution in revealing the reasoning from modern LLMs to users. Our study empirically demonstrates that this risk is real: well-intended features like ``show your reasoning'' can inadvertently erode the utilization of valuable information when it becomes available to humans.

\subsection{Limitations}

While our controlled experiment provides clear evidence of AI reasoning's effects on human trust and knowledge utilization, it has limitations that suggest future research.

First, our decision task is a simplified hiring scenario with a known ground truth formula. This abstraction allows us to isolate unavailable human knowledge (UHK) by withholding one feature from the AI (Personality Fit), but real-world decisions are often more complex. Follow-up studies should examine whether our findings generalize to expert decision-makers in real-world contexts (e.g., doctors diagnosing patients). Domain experts might respond differently in two ways: they could be less susceptible to AI overreliance due to greater confidence in their own knowledge, or they could be more susceptible if the AI's reasoning employs authoritative-sounding jargon from their field. Understanding how expertise and task complexity moderate the effects of AI reasoning is an important next step.

Second, our work focuses on a one-shot advisory scenario where the AI provides its recommendation and reasoning, and the human makes a final decision. As AI assistants become more interactive, researchers should investigate how trust and knowledge integration dynamics unfold over multi-round interactions. Future experiments should explore scenarios where users can interrupt the AI's reasoning to ask, "Did you consider this factor I know?" or where the AI proactively solicits user insights before finalizing recommendations. However, such interactive systems may first require understanding whether users are aware of UHK, recognize the information asymmetry between themselves and the AI, and can effectively articulate their unavailable knowledge to the system.

In summary, effective human-AI collaboration requires more than simply making AI transparent. It demands careful design and further research to ensure that human knowledge and judgment are not only preserved but actively enhanced in the collaborative process.

\newpage
\bibliography{ref}

\appendix
\section{Appendix}
\setcounter{table}{0}
\renewcommand{\thetable}{A\arabic{table}}

\subsection{Descriptive Statistics}

Table \ref{tab:descriptive_stats} presents the descriptive statistics for the participant sample, for the three conditions ``Extensive'' ($n=252$), ``No Reasoning'' ($n=250$) and ``Brief'' ($n=250$). 

The primary purpose of this table is to serve as a balance test, confirming that our randomization procedure resulted in comparable groups prior to the experimental manipulation. We report distributions for key demographic variables, including age, gender (percent female), employment status, annual household income, and educational attainment (percent with a college degree or higher). For categorical variables, we report counts and percentages. 

Additionally, we report baseline means and standard deviations (in parentheses) for the two AI-related measures: the ``AI Knowledge Test Score'' (0-5) and the ``AI Trust Score'' (0-50) by adding the scores of the related questions.

The final column displays the $p$-values from statistical tests comparing the three groups ($\chi^2$ tests for categorical variables and one-way ANOVA-tests for continuous variables). As shown, all $p$-values are well above conventional significance thresholds (e.g., all $p > 0.17$), indicating no statistically significant pre-existing differences between the experimental groups. This successful randomization supports the attribution of any subsequently observed differences in outcomes to our experimental treatment.

\begin{table}[htbp]
\centering
\caption{Descriptive Statistics and Balance Tests}
\label{tab:descriptive_stats}
\begin{tabular}{l c c c c}
\hline\hline
Treatment & Extensive & No  & Brief & p-value \\
 & (n=252) & (n=250) & (n=250) & \\
\hline
\multicolumn{5}{l}{\textit{Demographics}} \\
Age Distribution &  &  &  & 0.1726 \\
\quad 18-24 & 17 (6.7\%) & 13 (5.2\%) & 19 (7.6\%) &  \\
\quad 25-34 & 74 (29.4\%) & 84 (33.6\%) & 83 (33.2\%) &  \\
\quad 35-44 & 71 (28.2\%) & 73 (29.2\%) & 65 (26.0\%) &  \\
\quad 45-54 & 48 (19.0\%) & 34 (13.6\%) & 51 (20.4\%) &  \\
\quad 55-64 & 26 (10.3\%) & 38 (15.2\%) & 22 (8.8\%) &  \\
\quad 65 or older & 16 (6.3\%) & 8 (3.2\%) & 10 (4.0\%) &  \\
Female (\%) & 53.6 & 52.4 & 48.0 & 0.4620 \\
Hiring Experience (Yes, \%) & 65.9 & 58.4 & 58.4 & 0.1402 \\
Employment Status &  &  &  & 0.9500 \\
\quad Employed full-time & 140 (55.6\%) & 141 (56.4\%) & 139 (55.6\%) &  \\
\quad Employed part-time & 44 (17.5\%) & 39 (15.6\%) & 43 (17.2\%) &  \\
\quad Retired & 11 (4.4\%) & 10 (4.0\%) & 7 (2.8\%) &  \\
\quad Self-employed & 26 (10.3\%) & 28 (11.2\%) & 29 (11.6\%) &  \\
\quad Student & 6 (2.4\%) & 4 (1.6\%) & 9 (3.6\%) &  \\
\quad Unemployed & 25 (9.9\%) & 28 (11.2\%) & 23 (9.2\%) &  \\
Annual Household Income &  &  &  & 0.1753 \\
\quad Under \$25,000 & 23 (9.1\%) & 33 (13.2\%) & 26 (10.4\%) &  \\
\quad \$25,000 - \$49,999 & 57 (22.6\%) & 33 (13.2\%) & 52 (20.8\%) &  \\
\quad \$50,000 - \$74,999 & 49 (19.4\%) & 65 (26.0\%) & 59 (23.6\%) &  \\
\quad \$75,000 - \$99,999 & 50 (19.8\%) & 40 (16.0\%) & 44 (17.6\%) &  \\
\quad \$100,000 - \$149,999 & 41 (16.3\%) & 48 (19.2\%) & 45 (18.0\%) &  \\
\quad \$150,000 or more & 32 (12.7\%) & 31 (12.4\%) & 24 (9.6\%) &  \\
College Degree or Above (\%) & 29.4 & 33.6 & 26.4 & 0.2090 \\
\multicolumn{5}{l}{\textit{AI Perception}} \\
AI Knowledge Test Score (0-5) & 4.66 (0.66) & 4.68 (0.59) & 4.64 (0.69) & 0.8565 \\
AI Trust Score (0-50) & 30.89 (4.94) & 30.27 (5.43) & 30.30 (5.36) & 0.3237 \\
\hline\hline
\end{tabular}
\begin{flushleft}
\small
\textit{Note:} Values in parentheses are standard deviations. 
Significance levels: *** p$<$0.001, ** p$<$0.01, * p$<$0.05.
\end{flushleft}
\end{table}

\subsection{Experimental Interface}

Our experiment interface is carefully designed to isolate the effects of AI reasoning and control for potential confounding effects that may stem from the manipulation of reasoning length. 

First, to prevent rushed judgments by participants, each decision stage includes a 5-second countdown timer which disables decision buttons until adequate time has passed. This effectively disincentivizes the participants from exploiting our bonus payment scheme. 

Second, our AI reasoning processes are presented through a word-by-word streaming animation at 20 words per second. This streaming approach creates a sense of AI ``deliberation'' that mimics the real-time reasoning feature available on major LLM products (such as DeepSeek, ChatGPT, Gemini, etc.). Note that participants must wait until the AI has arrived at a conclusion before making a hiring decision. This requirement creates different timing across no reasoning and reasoning treatment groups (brief and extensive), which could affect the final decision accuracy. To control for the time the participant must wait before making the decision, we create a loading animation (i.e., a spinner) that indicates ``AI is thinking'' such that the waiting time will match the time in the brief reasoning group.

\begin{figure}[p]
    \centering
    \includegraphics[width=\textwidth]{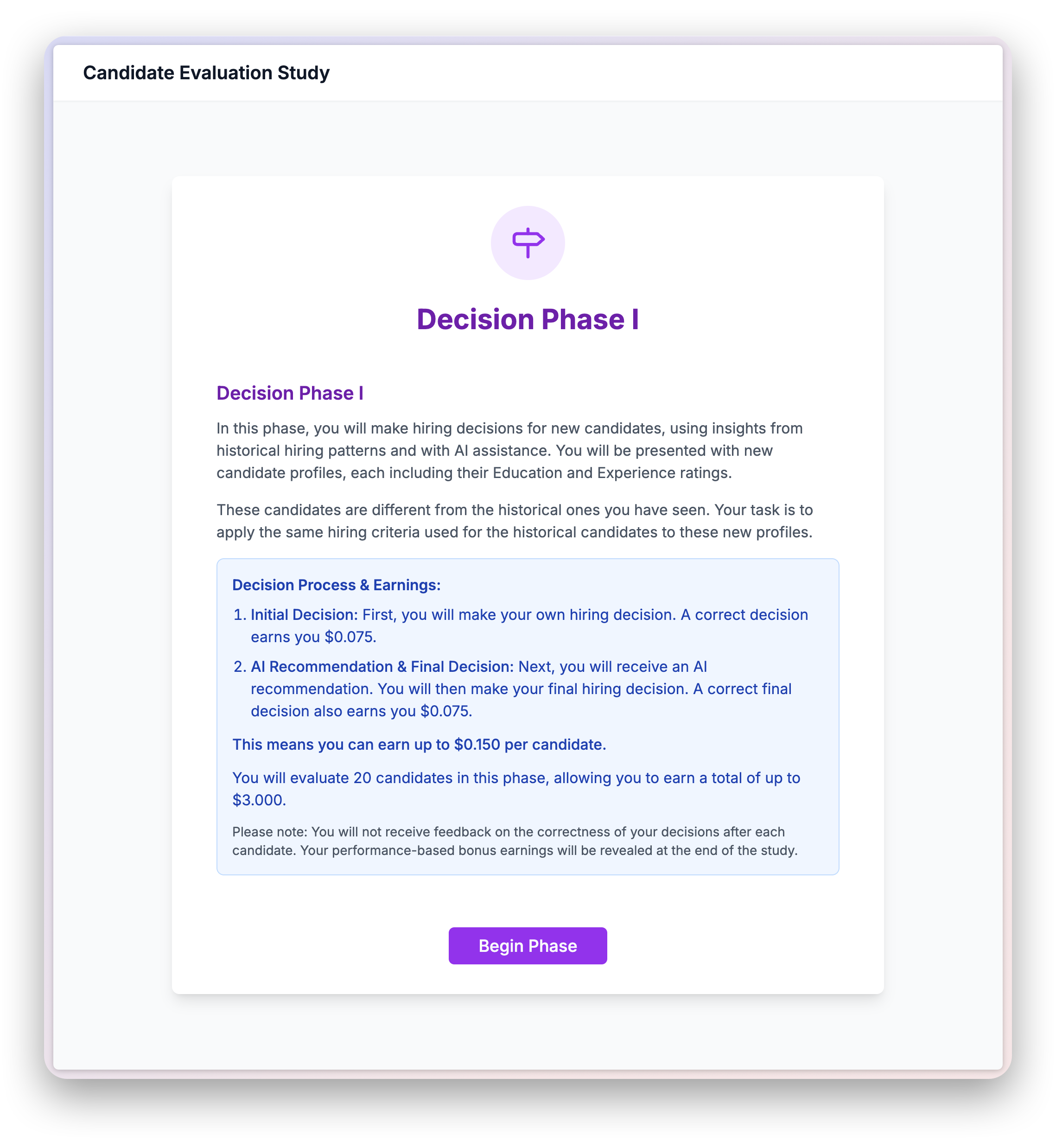}
    \caption{Phase I — Decision: Message view}
\end{figure}

\begin{figure}[p]
    \centering
    \includegraphics[width=\textwidth]{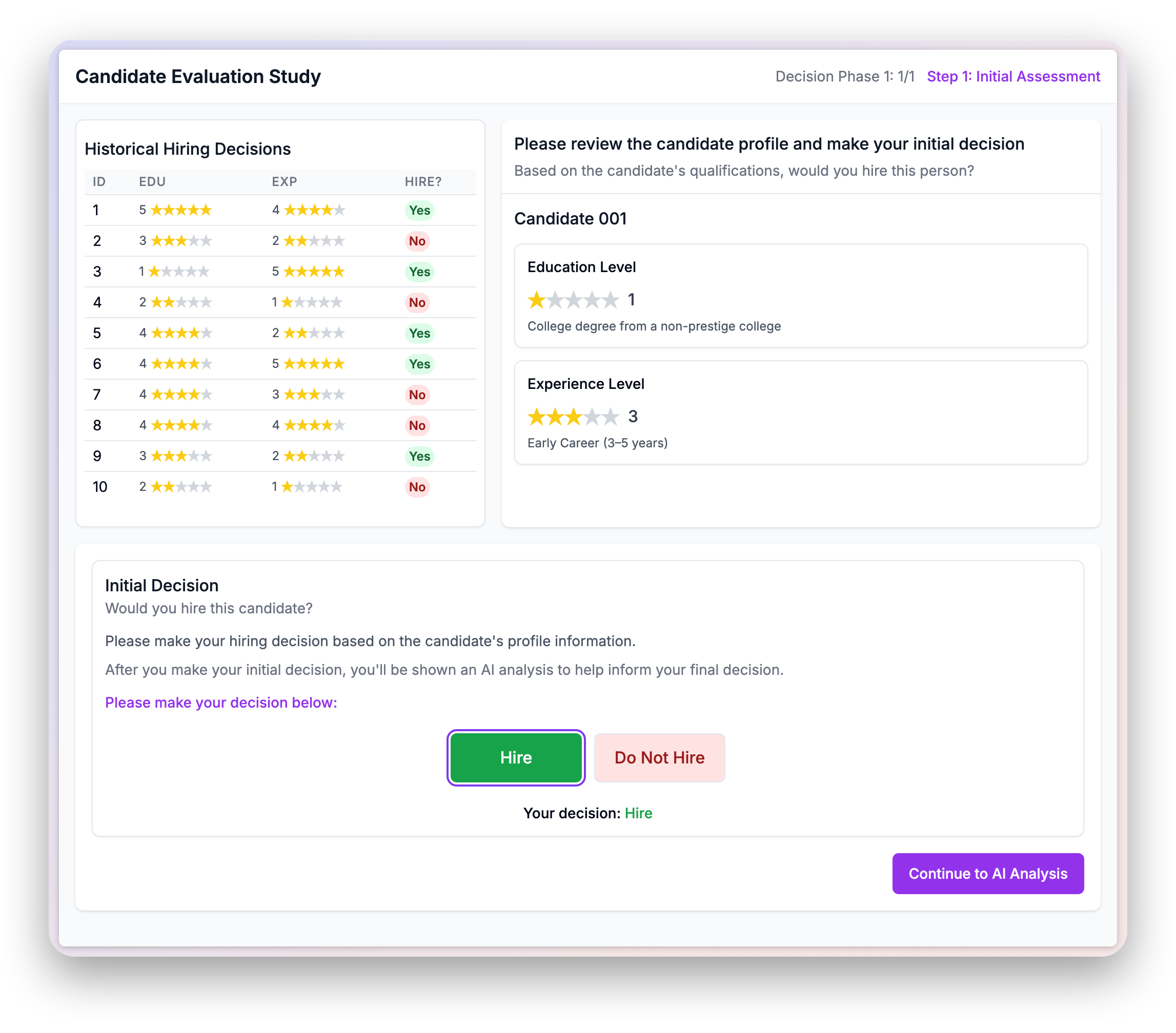}
    \caption{Phase I — Decision: Candidate initialization}
\end{figure}

\begin{figure}[p]
    \centering
    \includegraphics[width=\textwidth]{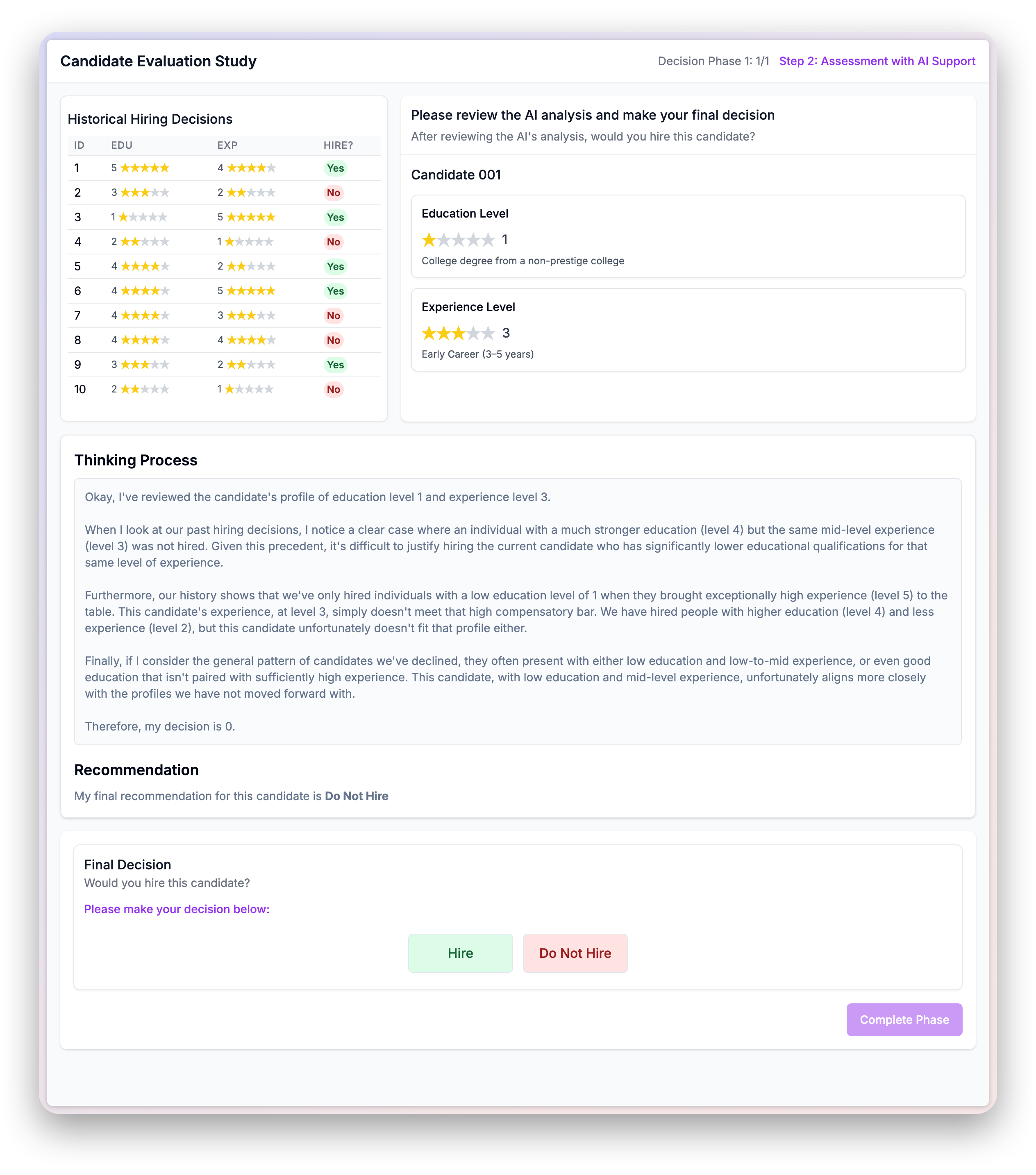}
    \caption{Phase I — Decision: Candidate reasoning (finish)}
\end{figure}

\begin{figure}[p]
    \centering
    \includegraphics[width=\textwidth]{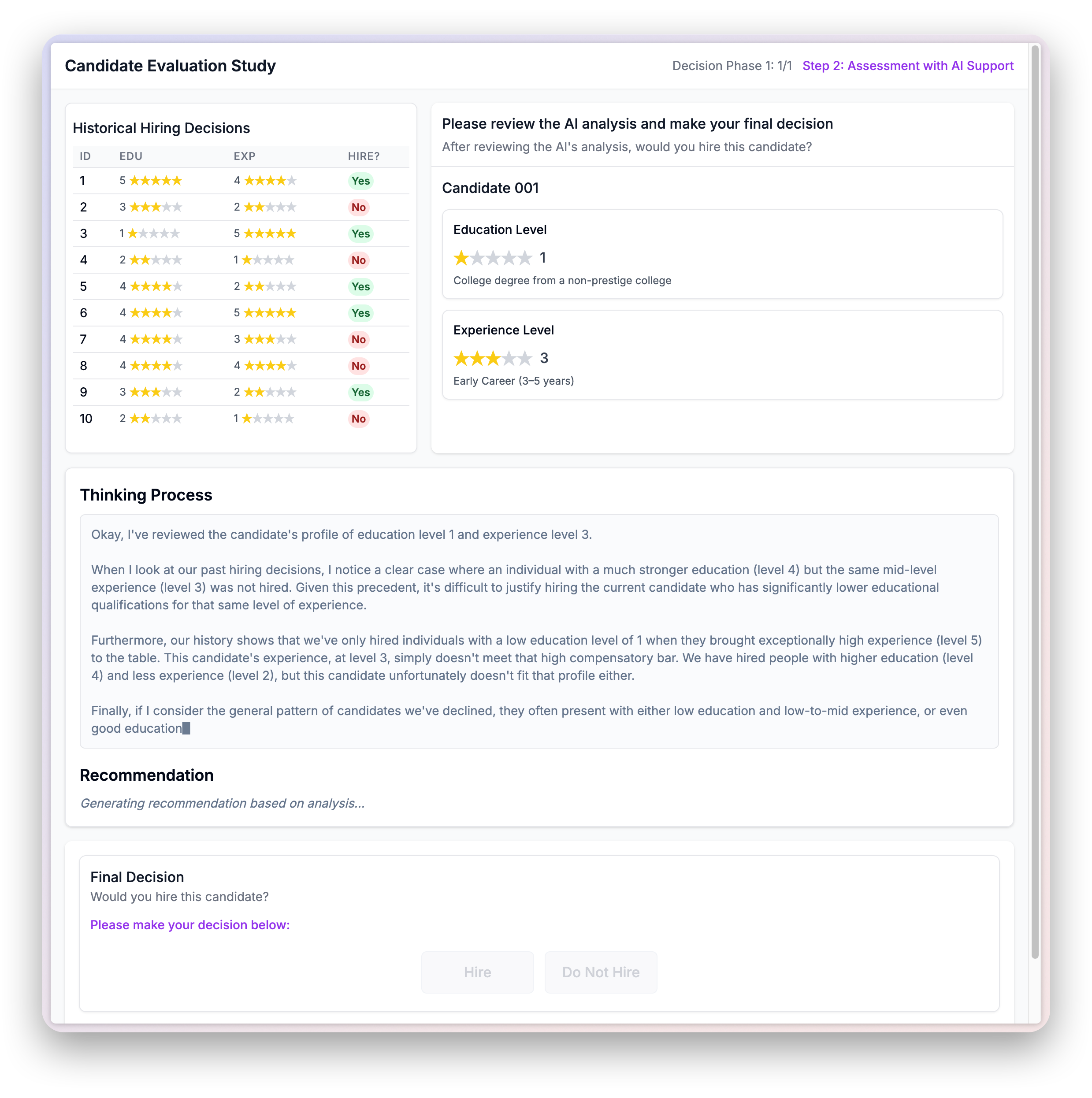}
    \caption{Phase I — Decision: Candidate reasoning (longer)}
\end{figure}

\begin{figure}[p]
    \centering
    \includegraphics[width=\textwidth]{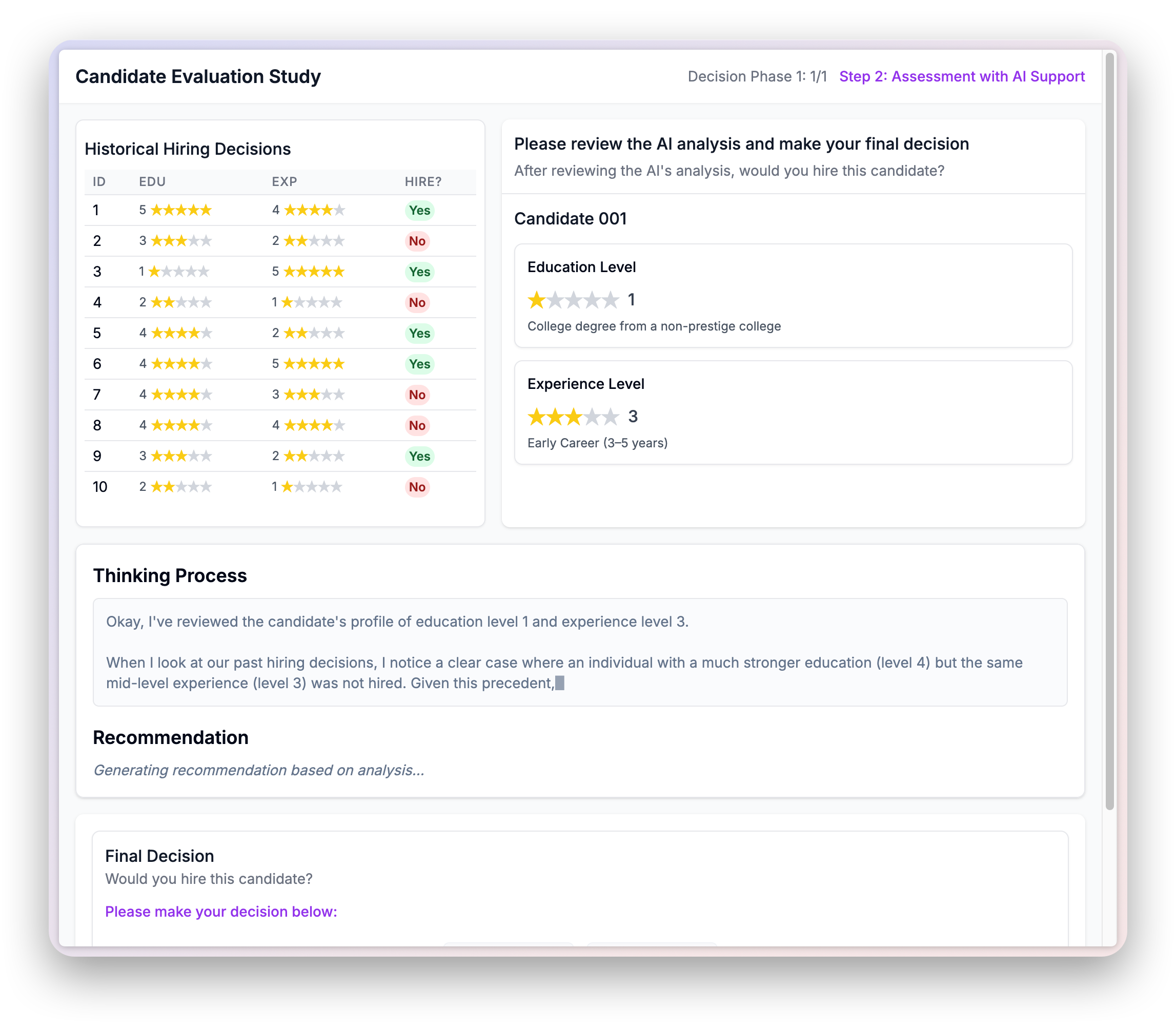}
    \caption{Phase I — Decision: Candidate reasoning}
\end{figure}

\begin{figure}[p]
    \centering
    \includegraphics[width=\textwidth]{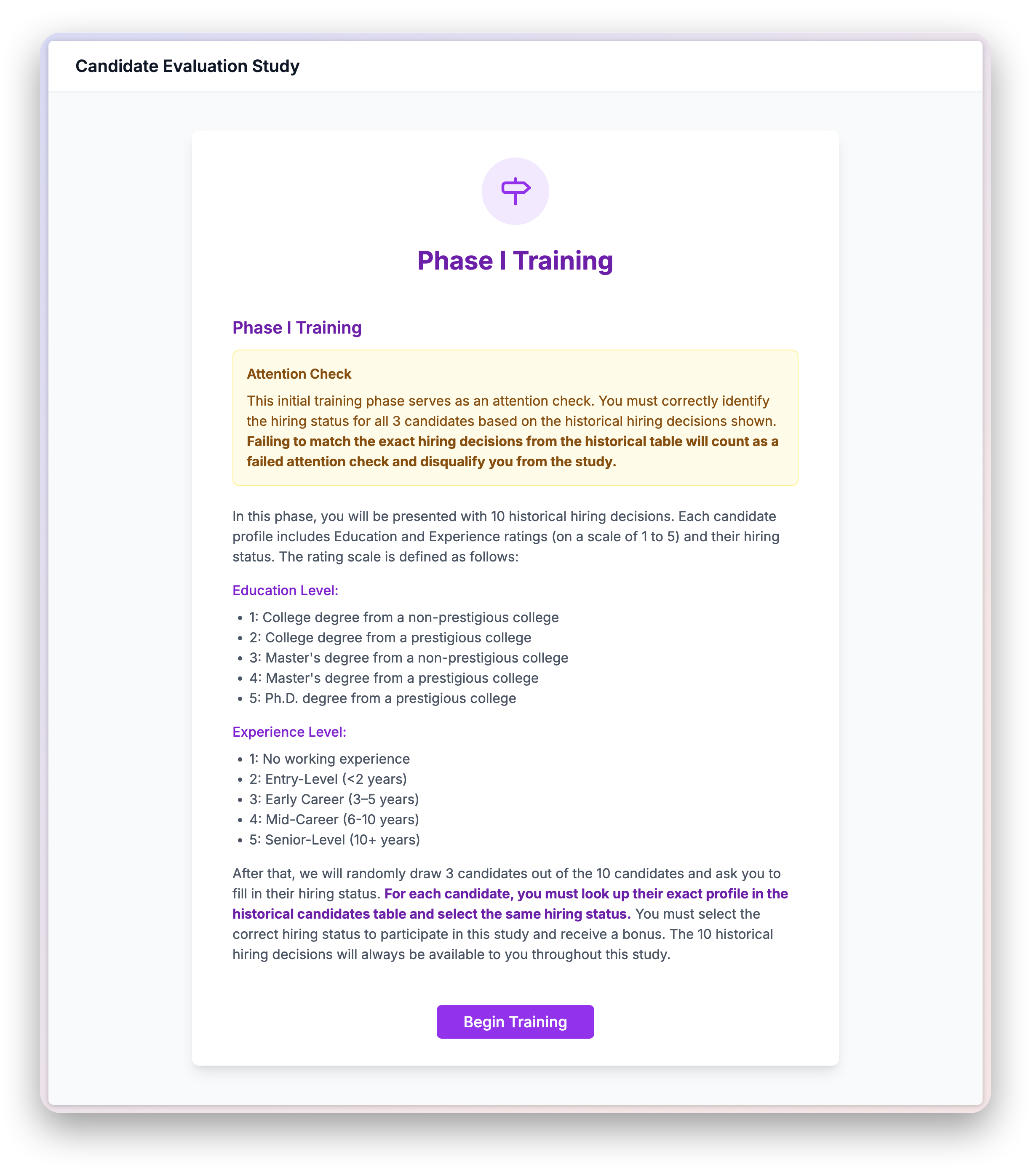}
    \caption{Phase I — Training: Message view}
\end{figure}

\begin{figure}[p]
    \centering
    \includegraphics[width=\textwidth]{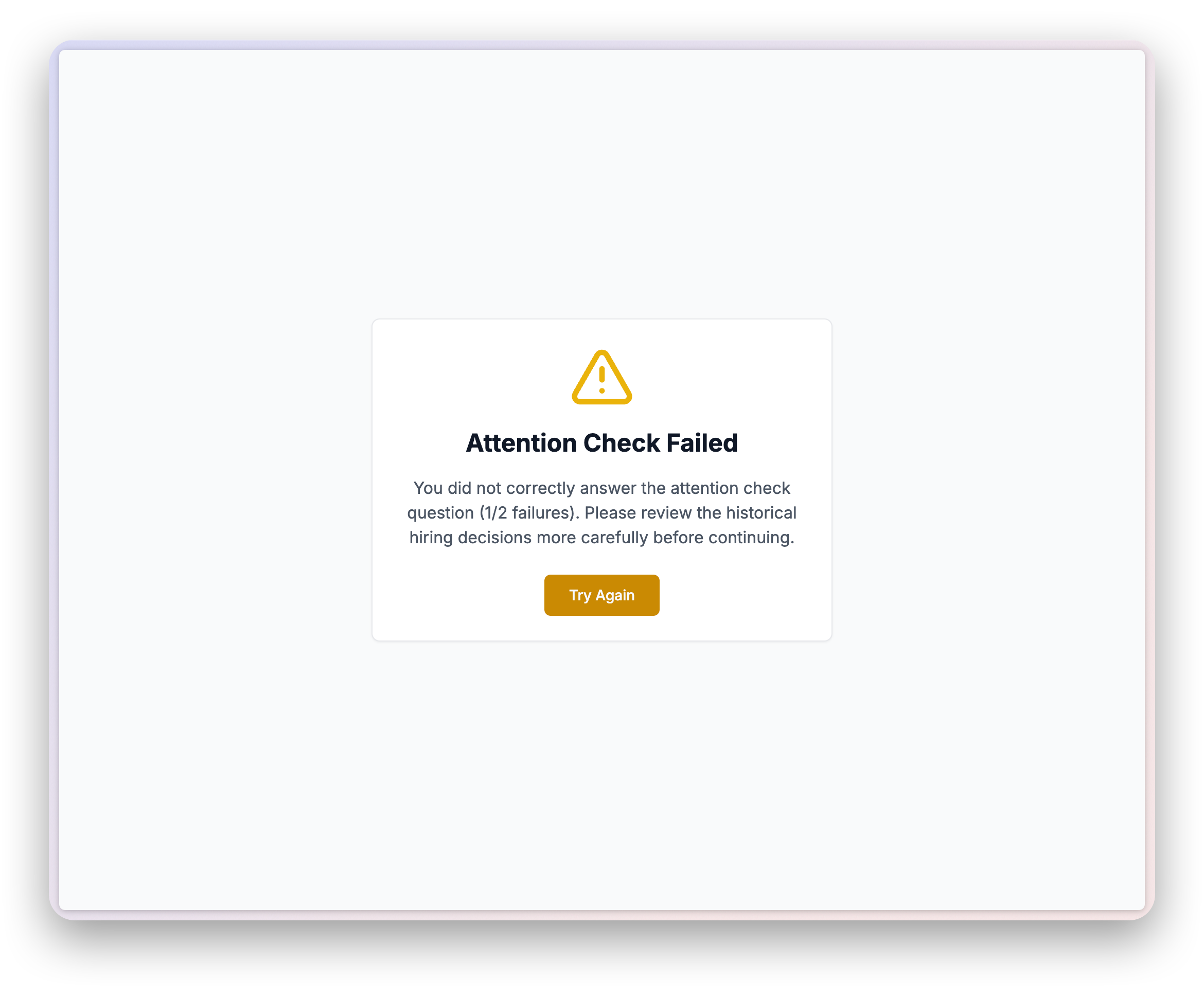}
    \caption{Phase I — Training: Attention check failed}
\end{figure}

\begin{figure}[p]
    \centering
    \includegraphics[width=\textwidth]{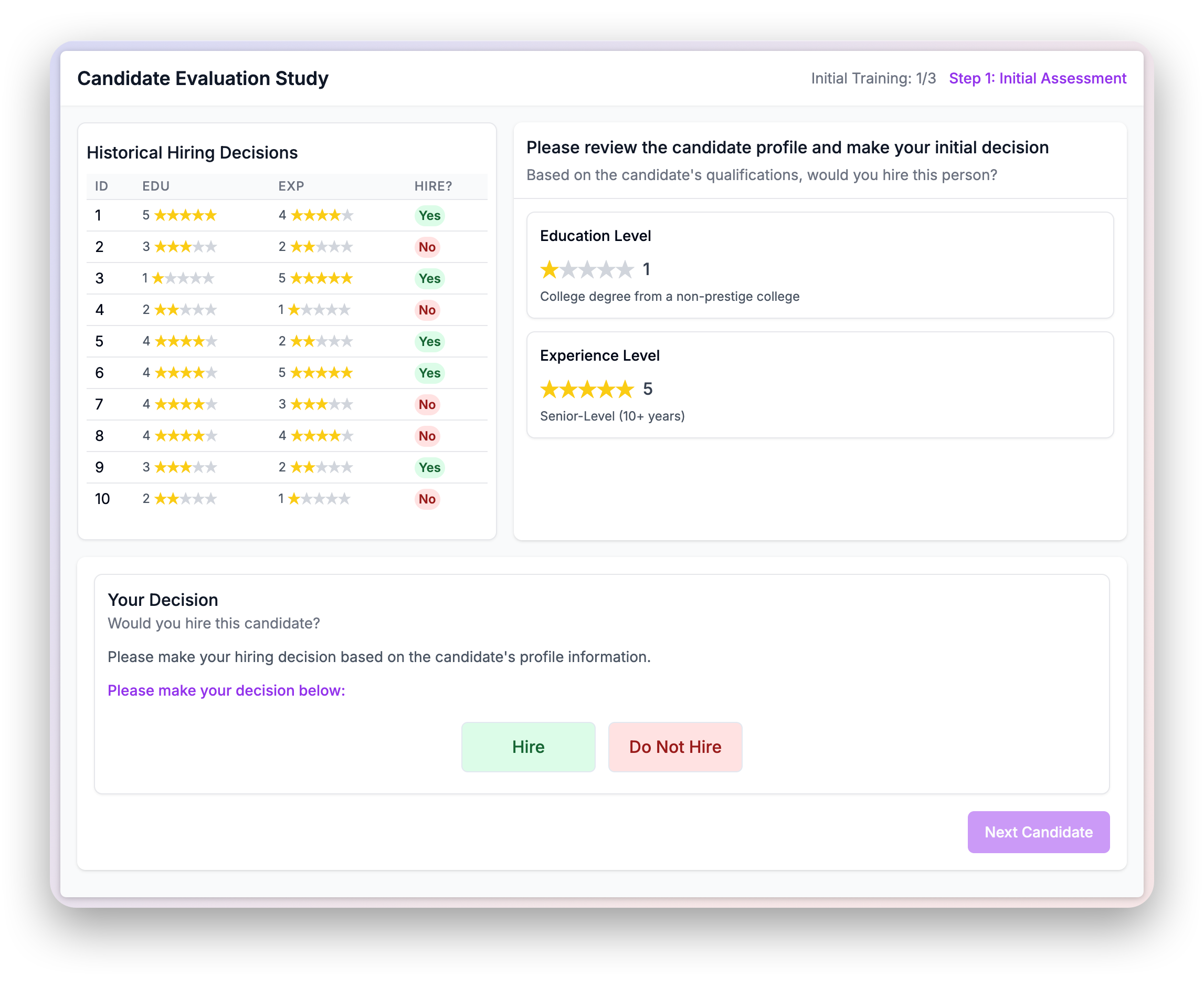}
    \caption{Phase I — Training: Candidate view}
\end{figure}

\begin{figure}[p]
    \centering
    \includegraphics[width=\textwidth]{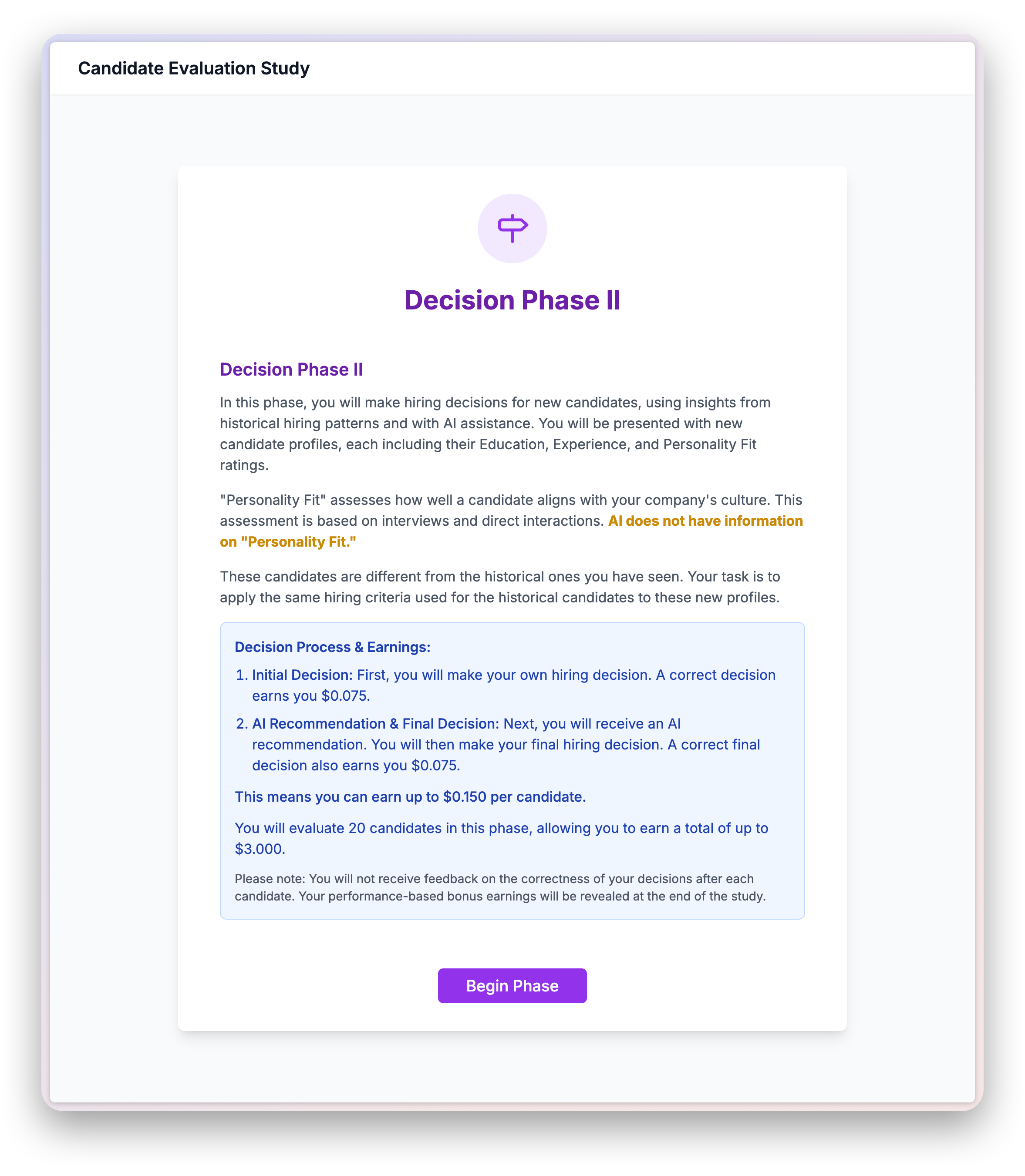}
    \caption{Phase II — Decision: Message view}
\end{figure}

\begin{figure}[p]
    \centering
    \includegraphics[width=\textwidth]{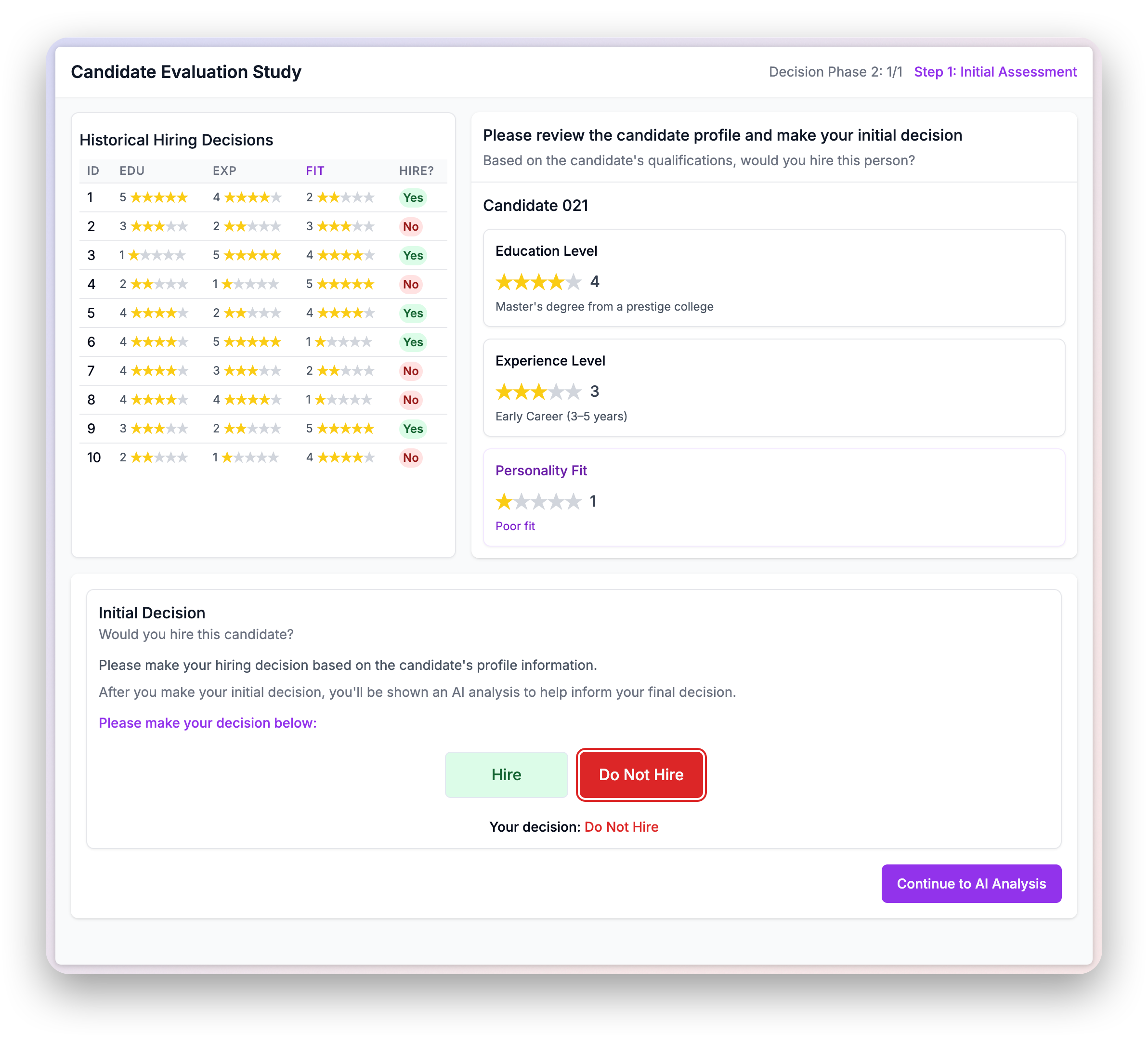}
    \caption{Phase II — Decision: Candidate initialization}
\end{figure}

\begin{figure}[p]
    \centering
    \includegraphics[width=\textwidth]{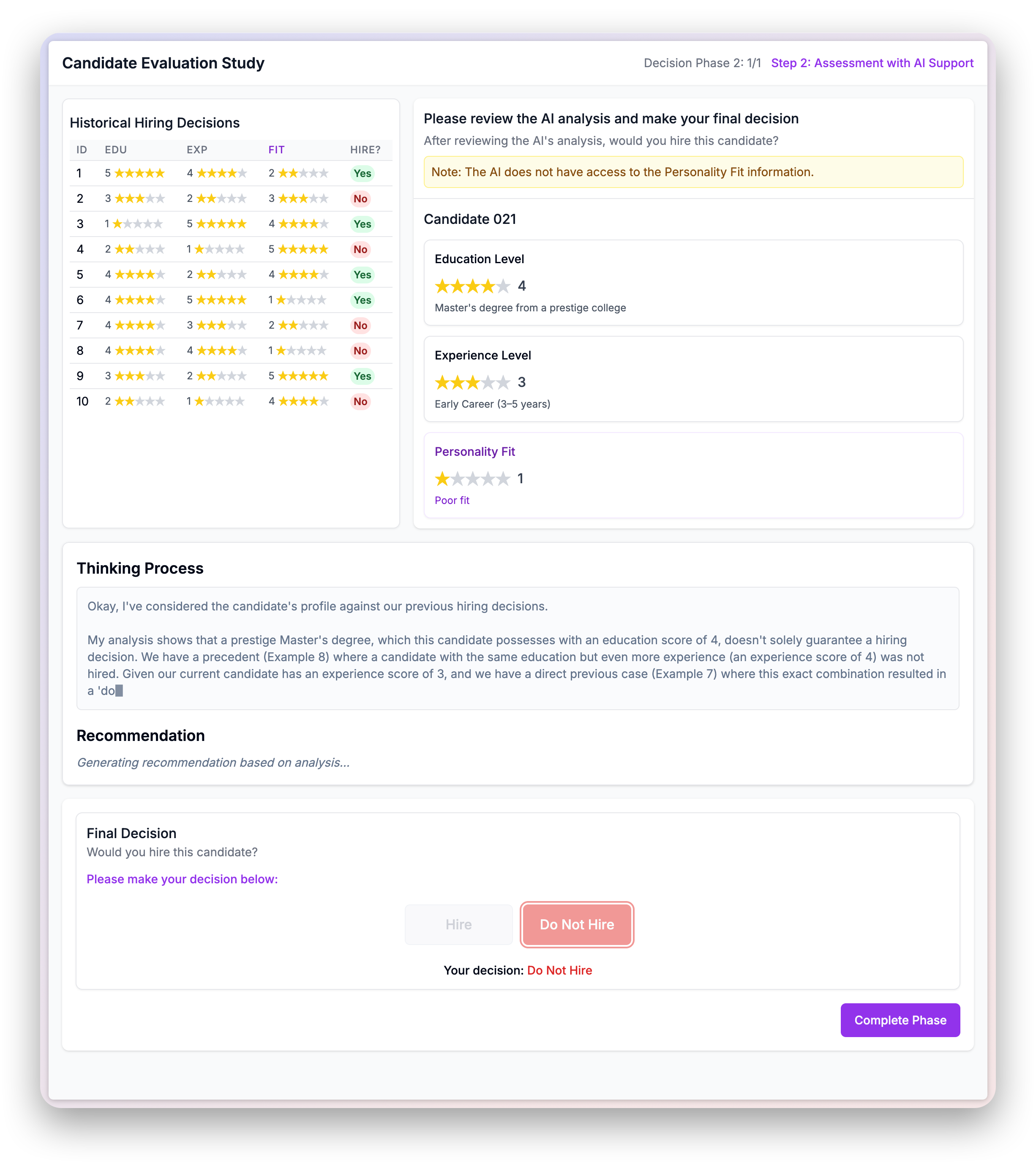}
    \caption{Phase II — Decision: Candidate reasoning}
\end{figure}

\begin{figure}[p]
    \centering
    \includegraphics[width=\textwidth]{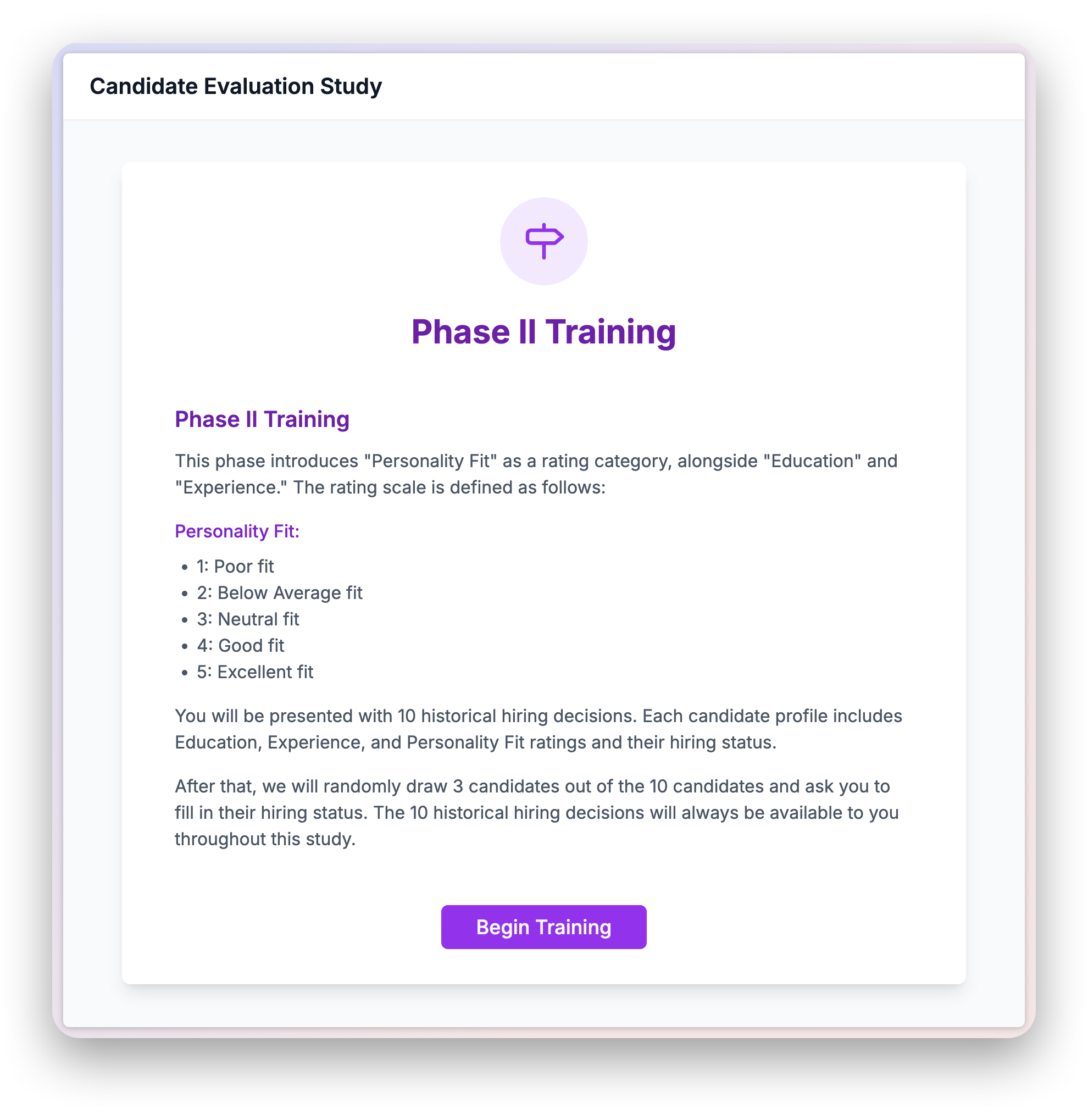}
    \caption{Phase II — Training: Message view}
\end{figure}

\begin{figure}[p]
    \centering
    \includegraphics[width=\textwidth]{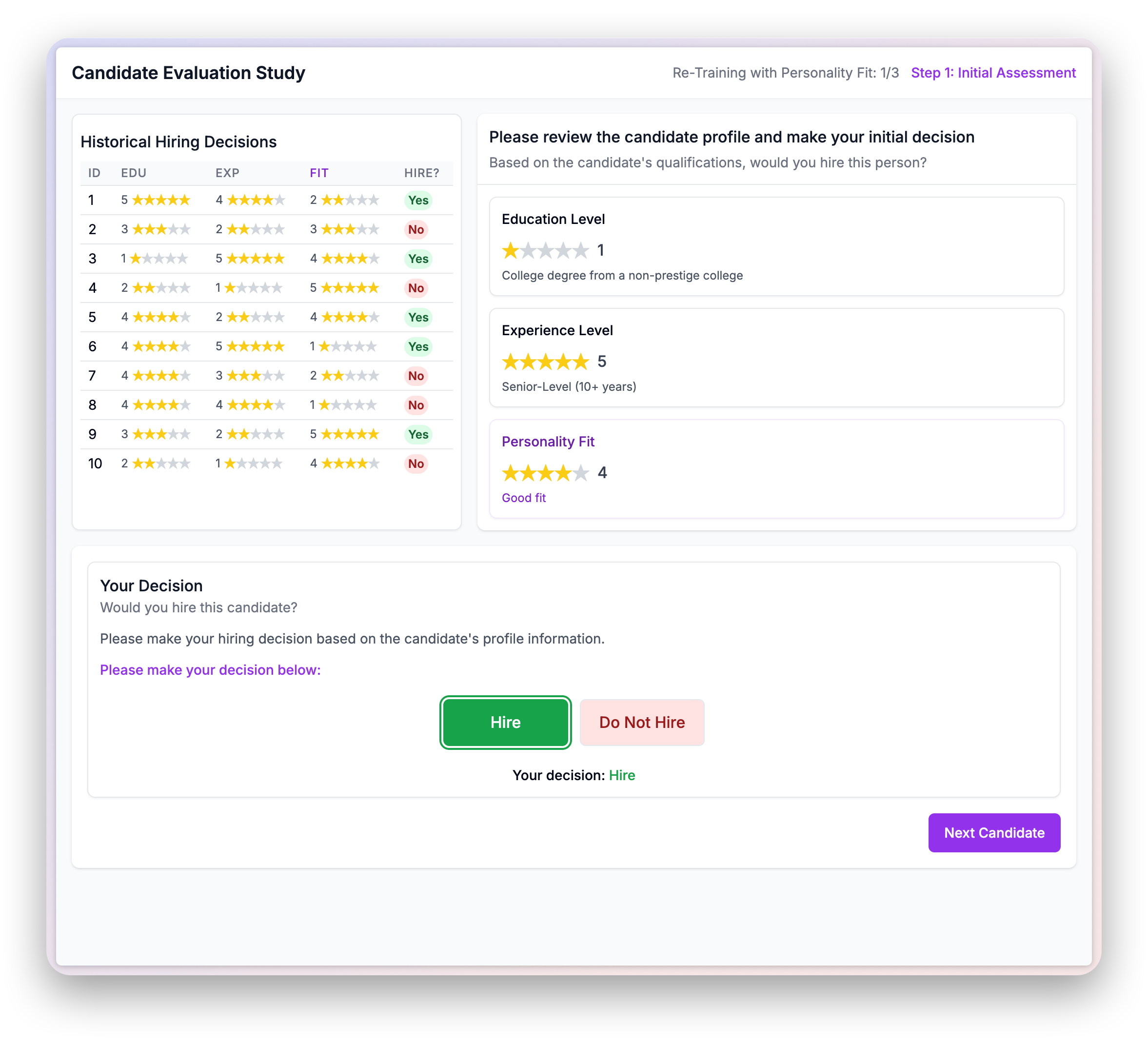}
    \caption{Phase II — Training: Candidate view}
\end{figure}

\pagebreak

\subsection{Instructions \& Surveys}

\subsubsection*{Instructions}

\textbf{Welcome to this Experiment on AI-Assisted Hiring Decisions}

\textbf{Your Role:} You will act as a Hiring Manager. Your primary task is to review information about job candidates and decide whether to hire them for a position based on their presented qualifications.

\textbf{Experiment Structure:} This experiment is divided into two main parts:
\begin{itemize}
    \item \textbf{Initial Survey:} First, you will complete a brief survey asking about your demographics and any previous experience you have with Artificial Intelligence (AI).
    \item \textbf{Hiring Task:} Following the survey, you will engage in a hiring task.
\end{itemize}

\textbf{Details of the Hiring Task:} The hiring task consists of two phases (Phase I and Phase II). Each phase follows the same structure:
\begin{itemize}
    \item \textbf{Training:} This will prepare you for the decision-making tasks by providing relevant information.
    \item \textbf{Actual Decision-Making:} This is where you will evaluate candidates and make hiring decisions.
\end{itemize}

\textbf{Total Earnings for the Experiment:} At the end of the experiment, we will calculate and reveal your total earnings. Your total earnings will consist of:
\begin{itemize}
    \item A \$4.00 base payment for completion of this study.
    \item Plus the performance bonuses you earned in Phase I (up to \$3.00).
    \item Plus the performance bonuses you earned in Phase II (up to \$3.00).
\end{itemize}
Therefore, you can earn a maximum total of \$10.00 in this experiment (\$4.00 base + \$3.00 Phase I bonus + \$3.00 Phase II bonus).

\newpage

\subsubsection*{Demographics}

\begin{enumerate} 
    \item \textbf{What is your age?}
    \begin{itemize} 
        \item Under 18
        \item 18-24
        \item 25-34
        \item 35-44
        \item 45-54
        \item 55-64
        \item 65 or older
    \end{itemize}
    \vspace{\baselineskip}

    \item \textbf{What is your gender?}
    \begin{itemize} 
        \item Male
        \item Female
        \item Non-binary/Other
        \item Prefer not to say
    \end{itemize}
    \vspace{\baselineskip}

    \item \textbf{What is the highest level of education you have completed?}
    \begin{itemize} 
        \item Less than high school
        \item High school diploma or equivalent
        \item Some college no degree
        \item Associate degree
        \item Bachelor's degree
        \item Graduate or professional degree
    \end{itemize}
    \vspace{\baselineskip}

    \item \textbf{What is your employment status?}
    \begin{itemize} 
        \item Employed full-time
        \item Employed part-time
        \item Self-employed
        \item Unemployed
        \item Student
        \item Retired
    \end{itemize}
    \vspace{\baselineskip}

    \item \textbf{What is your annual household income?}
    \begin{itemize} 
        \item Under \$25,000
        \item \$25,000 - \$49,999
        \item \$50,000 - \$74,999
        \item \$75,000 - \$99,999
        \item \$100,000 - \$149,999
        \item \$150,000 or more
    \end{itemize}
    \vspace{\baselineskip}

    \item \textbf{Do you have previous experience as a hiring manager or other type of hiring experiences?}
    \begin{itemize} 
        \item Yes
        \item No
    \end{itemize}
\end{enumerate}

\newpage

\subsubsection*{AI Perceptions and Knowledge}

\begin{enumerate} 
    \item \textbf{How familiar are you with the concept of artificial intelligence (AI)?}
    \begin{itemize} 
        \item Not at all familiar
        \item Slightly familiar
        \item Moderately familiar
        \item Very familiar
        \item Extremely familiar
    \end{itemize}
    \vspace{\baselineskip}

    \item \textbf{Have you ever used an AI-powered tool (such as ChatGPT, image generators, or voice assistants) in your daily life?}
    \begin{itemize} 
        \item Yes
        \item No
    \end{itemize}
    \vspace{\baselineskip}

    \item \textbf{What is your primary source of information about AI (can select multiple choices)?}
    \begin{itemize} 
        \item News outlets and magazines
        \item Social media platforms
        \item Formal education (courses, workshops, seminars)
        \item Online tutorials and blogs
        \item I don't actively seek information on AI
    \end{itemize}
    \vspace{\baselineskip}

    \item \textbf{Which of the following best describes artificial intelligence?}
    \begin{itemize} 
        \item A set of fixed computer programs that perform routine calculations
        \item Systems that mimic human cognitive functions---such as learning, reasoning, and problem solving
        \item Software that simply follows pre-defined rules without adaptation
        \item A tool used only for automating repetitive tasks
    \end{itemize}
    \vspace{\baselineskip}

    \item \textbf{This question is an attention check. What kind of decision do you need to make in this study?}
    \begin{itemize} 
        \item School Admission Decisions
        \item Layoff Decisions
        \item Hiring Decisions
        \item Inventory Prediction
    \end{itemize}
    \vspace{\baselineskip}

    \item \textbf{Which of the following applications is NOT typically powered by AI?}
    \begin{itemize} 
        \item Virtual assistants (e.g., Siri, Alexa)
        \item Recommendation systems on streaming platforms
        \item Autonomous vehicles
        \item Basic calculators
    \end{itemize}
    \vspace{\baselineskip}

    \item \textbf{Which statement best describes machine learning?}
    \begin{itemize} 
        \item A process where computers use statistical methods to improve performance over time through experience
        \item A method that relies solely on fixed algorithms with no change over time
        \item Random guessing without using data
        \item A technique for programming computers with hard-coded instructions
    \end{itemize}
    \vspace{\baselineskip}

    \item \textbf{What is the primary purpose of training data in machine learning?}
    \begin{itemize} 
        \item To test the final performance of the model
        \item To allow the model to learn patterns and improve its ability to make predictions
        \item To serve as a backup in case the model fails
        \item To manually program every decision the model makes
    \end{itemize}
    \vspace{\baselineskip}

    \item \textbf{What does ``overfitting'' mean in the context of machine learning?}
    \begin{itemize} 
        \item When a model fails to learn the training data
        \item When a model learns the training data too well, including its noise, and performs poorly on new data
        \item When a model perfectly predicts new data
        \item When a model has too few parameters to capture the data complexity
    \end{itemize}
\end{enumerate}

\newpage

\subsubsection*{Trust in AI}

\textbf{How much do you agree with the following statement?}
\vspace{1em}

The following items were rated on a 5-point Likert scale: \\ (1) \textit{Strongly Disagree}, (2) \textit{Disagree}, (3) \textit{Neither Agree nor Disagree}, (4) \textit{Agree}, (5) \textit{Strongly Agree}.

\begin{enumerate} 
    \item \textit{``I trust that AI systems are capable of making accurate decisions.''}
    \item \textit{``I prefer decisions made by human experts over those made by AI, even if the AI is more statistically accurate.''}
    \item \textit{``I believe that AI systems are more objective than humans in their decision-making.''}
    \item \textit{``I feel comfortable delegating routine tasks to AI systems.''}
    \item \textit{``When an AI system makes a mistake, I lose trust in its overall performance.''}
    \item \textit{``For complex or emotional decisions, I believe human judgment is essential.''}
    \item \textit{``I am more willing to use an AI system if I have some control over its outputs.''}
    \item \textit{``I worry that AI systems are likely to be biased in their decisions.''}
    \item \textit{``I believe AI systems perform better than humans when it comes to routine, low-risk tasks.''}
    \item \textit{``Overall, I feel comfortable relying on AI to assist me with decision-making.''}
\end{enumerate}
\vspace{\baselineskip}

\textbf{This question is an attention check. How many phases are there in the hiring decision tasks?}
\begin{itemize} 
    \item 1
    \item 2
    \item 3
    \item 4
\end{itemize}

\subsection{Prompts}\label{app:prompts}

Here we include the prompt template we used for querying the large language model in Table \ref{tab:prompt}. 

\begin{table}[h!]
    \centering
    \resizebox{!}{0.6\textwidth}{
        \begin{tabular}{|p{\textwidth}|}
        \hline
        \textbf{You are a helpful assistant whose goal is to decide whether the job candidate should be hired.} \\
        \textbf{Instructions: }
        You will receive: \\
        \textbf{Previous Examples} \\
        You will be given previous examples of hiring decisions, which includes two features and one outcome:
        \begin{itemize}
            \item \textbf{Education}: this is a score from 1 to 5, where 1 means a Bachelor's degree from a non-prestige college, 2 means a Bachelor’s degree from a prestige college, 3 means a Master’s degree from a non-prestige college, 4 means a Master’s degree from a prestige college, and 5 means a Ph.D. degree from a prestige college. no education and 3 means a PhD from a top university.
            \item \textbf{Experience}: this is a score from 1 to 5, where 1 means no experience, 2 means less than 2 years of experience, 3 means more than 2 years but less than 5 years, 4 means more than 5 years but less than 10 years, and 5 means more than 10 years of experience. and 3 means 10+ years of experience in a relevant role.
            \item \textbf{Hiring outcome}: this is an outcome of 1 or 0, where 1 means hired and 0 means not hired.
        \end{itemize}
        \textbf{Candidate Description} \\
        You will also be given candidate's qualifications, which includes:
        \begin{itemize}
            \item \textbf{Education}: this is a score from 1 to 5, where 1 means a Bachelor's degree from a non-prestige college, 2 means a Bachelor’s degree from a prestige college, 3 means a Master’s degree from a non-prestige college, 4 means a Master’s degree from a prestige college, and 5 means a Ph.D. degree from a prestige college.
            \item \textbf{Experience}: this is a score from 1 to 5, where 1 means no experience, 2 means less than 2 years of experience, 3 means more than 2 years but less than 5 years, 4 means more than 5 years but less than 10 years, and 5 means more than 10 years of experience.
        \end{itemize}
        Your task is to:
        Evaluate whether the candidate should be hired based on \textbf{experience} and \textbf{education}.
        Only output 1 or 0.
        \\
        \textbf{Previous Examples}: \\
        \{`education': 5, `experience': 4, `hire outcome': 1\} \\
\{`education': 3, `experience': 2, `hire outcome': 0\} \\
\{`education': 1, `experience': 5, `hire outcome': 1\} \\
\{`education': 2, `experience': 1, `hire outcome': 0\} \\
\{`education': 4, `experience': 2, `hire outcome': 1\} \\
\{`education': 4, `experience': 5, `hire outcome': 1\} \\
\{`education': 4, `experience': 3, `hire outcome': 0\} \\
\{`education': 4, `experience': 4, `hire outcome': 0\} \\
\{`education': 3, `experience': 2, `hire outcome': 1\} \\
\{`education': 2, `experience': 1, `hire outcome': 0\} \\
        \textbf{Candidate Description}: 
\{`education': 1, `experience': 3\}
        \\
        \textbf{Task} \\
        Hiring decision for the candidate: \\
        \hline
        \end{tabular}
    }
    \caption{Prompt Example}
    \label{tab:prompt}
\end{table}

\subsection{AI Reasoning}
\label{app: reasoning_prompt}

Extensive AI reasoning is derived from the full, raw reasoning generated by ``gemini-2.5-pro-preview'' through the official web interface provided by Google. As such, we use the official model parameters used by Google. Because the full reasoning is excessively long, we use the following prompt to generate reasoning text that can be read in a reasonable amount of time:

\begin{quote}
\begin{lstlisting}
{text}

---
Read the above reasoning process and provide a short, concise text with THREE alternative reasons that support the final decision. You must provide THREE reasons, each has its own paragraph. Please use a first-person tone, avoiding bullet points or phrases like `firstly' and `secondly.' Act as the decision-maker and state that decision at the end. The response should read like an ongoing narrative that walks through your thought process from the start-as if you are figuring things out in the moment-instead of just explaining a decision already made. In other words, you should not use past-tense.

Output the results directly, do not add anything before the results.
\end{lstlisting}
\end{quote}

Similarly, the brief AI reasoning is generated using the following prompt from the original full reasoning:

\begin{quote}
\begin{lstlisting}
{text}
---

Based on the text above, extract the first minimal-viable reasoning. You can only truncate the original text with slight modification for fluency. The output should be a single paragraph, and the decision should be at the end of the paragraph. The decision should be clear and unambiguous, and it should not be something like "difficult to decide". The output should not contain any bullet points or lists. The first sentence and the last sentence should be exactly the same as the original text.

It should read like a narrative, as opposed to one that justifies a pre-made decision (that is, at the start of the text, you should act as if you are at the beginning of the reasoning process, and no decisions has been made at that point).

--- example start ---
Examining the historical data, there is no exact match for this specific combination. However, comparing the candidate's profile to similar cases provides strong indicators. A particularly relevant example is candidate 5, where a candidate with significantly higher education but the same level of experience as the current candidate was not hired. Given that a profile with superior education but equivalent experience was rejected, it is highly improbable that a candidate with lower education and the same experience level would be deemed suitable based on the established patterns. The decision is 0, do not hire.
--- example end ---

Now follow this example and output your results. Output the results directly, do not add anything before the results.
\end{lstlisting}
\end{quote}

Table \ref{tab:reasoning_example} shows a typical reasoning process shown to the participants across three conditions. To easily parse the hiring decision, we prompted the model to output a numerical decision of 0 (meaning DOES NOT HIRE) and 1 (meaning HIRE).    

\begin{table}[h!]
\centering
\caption{AI Reasoning Example}
\label{tab:reasoning_example}
\renewcommand{\arraystretch}{1.5}
\begin{tabular}{>{\raggedright}p{2cm}  >{\raggedright}p{12cm} >{\centering\arraybackslash}p{1.5cm}}
\toprule
\textbf{Condition} & \textbf{Reasoning} & \textbf{Decision} \\
\midrule

\textbf{Extensive}  &
When I look at our past hiring decisions, I notice a clear case where an individual with a much stronger education (level 4) but the same mid-level experience (level 3) was not hired. Given this precedent, it's difficult to justify hiring the current candidate who has significantly lower educational qualifications for that same level of experience. \newline 
\newline 
 Furthermore, our history shows that we've only hired individuals with a low education level of 1 when they brought exceptionally high experience (level 5) to the table. This candidate's experience, at level 3, simply doesn't meet that high compensatory bar. We have hired people with higher education (level 4) and less experience (level 2), but this candidate unfortunately doesn't fit that profile either. \newline 
\newline 
 Finally, if I consider the general pattern of candidates we've declined, they often present with either low education and low-to-mid experience, or even good education that isn't paired with sufficiently high experience. This candidate, with low education and mid-level experience, unfortunately aligns more closely with the profiles we have not moved forward with.
 \newline 
\newline 
 Therefore, my decision is 0.
&
  Not Hire \\
\midrule

\textbf{Brief}  &
 When I look at our past hiring decisions, I notice a clear case where an individual with a much stronger education (level 4) but the same mid-level experience (level 3) was not hired. Given this precedent, it's difficult to justify hiring the current candidate who has significantly lower educational qualifications for that same level of experience. 
 \newline 
\newline 
Therefore, my decision is 0.&
  Not Hire \\
\midrule

\textbf{No} &
  --- &
  Not Hire \\
\bottomrule
\end{tabular}
\end{table}

\bibliographystyle{science/sciencemag}

\end{document}